\documentclass[preprint,3p,12pt]{elsarticle}
\usepackage{mathrsfs}
\usepackage{amsmath}
\usepackage{stmaryrd}
\usepackage{bbding}
\usepackage{dcolumn}
\usepackage{graphicx}
\usepackage{amsfonts}
\usepackage{amssymb}
\usepackage{psfrag}
\usepackage{wrapfig}
\usepackage{subfigure}
\usepackage{makeidx}
\usepackage{bm}
\usepackage{epsf}
\usepackage{epsfig}
\usepackage{setspace}
\usepackage{graphicx}
\usepackage{epstopdf}
\usepackage{psfrag}
\usepackage{subfigure}
\usepackage{booktabs}
\usepackage{color}
\usepackage{comment}
\begin{document}

\title{UGKWP method for polydisperse gas-solid particle multiphase flow}

\author[HKUST1]{Xiaojian Yang}
\ead{xyangbm@connect.ust.hk}

\author[HKUST1]{Wei Shyy}
\ead{weishyy@ust.hk}

\author[HKUST1,HKUST2,HKUST3]{Kun Xu\corref{cor1}}
\ead{makxu@ust.hk}	

\address[HKUST1]{Department of Mechanical and Aerospace Engineering, Hong Kong University of Science and Technology, Clear Water Bay, Kowloon, Hong Kong, China}
\address[HKUST2]{Department of Mathematics, Hong Kong University of Science and Technology, Clear Water Bay, Kowloon, Hong Kong, China}
\address[HKUST3]{Shenzhen Research Institute, Hong Kong University of Science and Technology, Shenzhen, China}
\cortext[cor1]{Corresponding author}

\begin{abstract}
The gas-particle flow with multiple dispersed solid phase is associated with complicated multiphase flow dynamics.
In this paper, a unified algorithm will be proposed for the study of gas-solid particle multiphase flow.
The gas-kinetic scheme (GKS) is used to simulate the continuum gas phase and the multiscale unified gas-kinetic wave-particle (UGKWP) method
is developed for the multiple dispersed solid particle phase.
At the same time, the momentum and energy exchanges between gas-particle phases will be included under the GKS-UGKWP framework.
For each disperse solid particle phase, the decomposition of deterministic wave and statistic particle in UGKWP is based on the local cell's Knudsen number.
The method for solid particle phase can become the Eulerian fluid approach at small cell's Knudsen number and
Lagrangian particle approach at large cell's Knudsen number.
This is very significant for simulating dispersed particle phases at different Knudsen numbers due to the variation of
physical properties in individual particle phase, such as the particle diameter, material density, and the corresponding mass fraction inside each control volume. For the gas phase, the GKS is basically an Eulerian approach for the NS solution.
Therefore, the GKS-UGKWP method for the gas-particle flow unifies the Eulerian-Eulerian (EE) and Eulerian-Lagrangian (EL) methods.
An optimal strategy can be obtained for the solid particle phase with the consideration of physical accuracy and numerical efficiency.
Two cases of gas-solid fluidization system, i.e., one circulating fluidized bed and one turbulent fluidized bed, are simulated. The typical flow structures of the fluidized particles are captured, and the time-averaged variables of the flow field agree well with the experimental measurements.
In addition, the shock particle-bed interaction is studied by the proposed GKS-UGKWP, which validates the method for polydisperse gas-particle system in the supersonic case, where the dynamic evolution process of the particle cloud is investigated.
\end{abstract}


\maketitle

\section{Introduction}
Gas-particle two-phase flow appears in chemical, petroleum, environmental and other industries.
Quantitative studies of the system is of great importance in both basic scientific research and practical industrial production \cite{Gasparticle-review-Gewei-zhang2023numerical, Gasparticle-momentmethod-Fox2013computational}.
Most granular flow system includes multiple types of solid particles with different material densities, diameters, shapes, etc,
which is named as the polydisperse gas-particle flow.
For the polydisperse system, it may become problematic to regard all kinds of solid particles as a single phase,
especially as particle properties differ significantly from each other.
For instance, the rate of chemical reaction is highly dependent on the particle size, where the effect from
the particle size distribution (PSD) cannot be ignored \cite{Gasparticle-review-zhongwenqi-zhong2016cfd, Gasparticle-polydisperse-drag-EMMS-wangjunwu2019emms}.
Besides the interaction between gas flow and solid particles,
the interaction between solid and solid particles, which is modeled by the so-called solid-to-solid drag in the community of gas-particle flow,
also plays a significant role in accurately predicting the particles' behavior \cite{Gasparticle-polydisperse-solid-drag-syamlal1987particle, Gasparticle-polydisperse-solid-drag-mathiesen1999experimental}.
Therefore, developing an advanced computational fluid dynamics (CFD) tool for polydisperse gas-particle flow is more challenging than its monodisperse counterparts due to the increased complexity in the multi-particle system \cite{Gasparticle-review-Gewei-zhang2023numerical}.

Generally two approaches, Eulerian-Eulerian (EE) approach and Eulerian-Lagrangian (EL) approach,
are used for the study of gas-particle two-phase flow.
In both approaches, the gas phase is described by the Navier-Stokes (NS) equations, i.e., the so-called Eulerian approach;
while the treatment of the solid particle phase can be Eulerian or Lagrangian ones.
In the EE approach, the particle phase is modeled  as a continuum fluid, and the hydrodynamic solvers are used
in the simulation \cite{Gasparticle-book-gidaspow1994multiphase, Gasparticle-Abgrall-saurel1999multiphase, Gasparticle-polydisperse-KTGF-huilin2003hydrodynamics}. In the EL approach, all solid particles or parcels,
standing for a group of solid particles with the same properties, are tracked individually by solving the Newtonian equation of particle motion. At the same time, the collisions between solid particles are modeled, such as these collision rules in the discrete element method (DEM) \cite{Gasparticle-DEM-tsuji1993discrete, Gasparticle-polydisperse-drag-compare-zhang2017assessment} and in the multiphase particle-in-cell method (MP-PIC) \cite{Gasparticle-PIC-snider2001incompressible, Gasparticle-PIC-snider2007three, Gasparticle-PIC-novel-verma2020novel}.
In general, the EL approach works well in all flow regimes. Although the EE approach may not be able to accurately predict the particular flow when the Knudsen number (\text{Kn}) of the solid phase is large, it is still the dominant method in practical engineering applications
due to its computational efficiency \cite{Gasparticle-review-zhongwenqi-zhong2016cfd, Gasparticle-review-Gewei-zhang2023numerical}.
In addition to the aforementioned methods, other commonly-used numerical methods for granular flow include, but are not limited to, method of moment (MOM) \cite{Gasparticle-momentmethod-Fox2013computational, Gasparticle-polydisperse-KTGF-fox-fan2008segregation}, direct simulation of Monte Carlo (DSMC) \cite{Gasparticle-DSMC-ZhaoHB-he2015differentially}, compressible multiphase particle-in-cell method (CMP-PIC) \cite{Gasparticle-PIC-DEM-tian2020compressible}, and hybrid EL method \cite{Gasparticle-hybrid-EL-qinghong-luhuilin-2017coupled}, and many others.

For the solid particle evolution, the dynamics from the particle free transport with the interaction with the gas phase
and inter-particle collisions should be modeled \cite{Gasparticle-momentmethod-Fox2013computational, WP-gasparticle-fluidized-yang2023unified}.
For the polydisperse flow, the inter-particle collision includes the monodisperse and polydisperse types of particles.
For the EL approach, such as DEM, the effect on the particle in the polydisperse case can be added straightforwardly in the simulation,
since all solid particles' transport will be tracked with the explicit inter-particle collision according to the collision law \cite{Gasparticle-polydisperse-DEM-YuAB-feng2004discrete, Gasparticle-polydisperse-drag-compare-zhang2017assessment}.
However, the computational cost is high, especially at small \text{Kn}, due to the tracking of tremendous amount of particles.
For the EE approach, the multi-fluid strategy is one of the most commonly adopted methods in the polydisperse flow study,
where many sets of governing equations are employed to describe the disperse phases \cite{Gasparticle-polydisperse-solid-drag-mathiesen1999experimental, Gasparticle-polydisperse-KTGF-huilin2003hydrodynamics, Gasparticle-polydisperse-drag-EMMS-wangjunwu2019emms}.
For example, in the study of hydrodynamic behaviors from PSD, the particle phase is modeled in different discrete phases
according to the particle size \cite{Gasparticle-polydisperse-drag-EMMS-wangjunwu2019emms}.
In the multi-fluid model, the closure of one solid phase will involve the properties of other disperse phases.
One of the important factors is about the determination of the solid-to-solid drag in the momentum exchange between different disperse phases
\cite{Gasparticle-polydisperse-solid-drag-syamlal1987particle, Gasparticle-polydisperse-solid-drag-mathiesen1999experimental, Gasparticle-polydisperse-KTGF-fox-fan2008segregation}.
Another approach is to use one set of governing equations for the whole solid phase, and additional modifications are added
with the consideration of different particle sizes \cite{Gasparticle-polydisperse-KTGF-PBM-emms-chen2013coarse}.
In addition, the gas-solid interaction also plays a significant role in monodisperse/polydisperse particular flow in both EE and EL approaches.
In particular, the drag on solid particle from the gas flow is of great importance to accurately predict the flow field of the gas-particle system \cite{Gasparticle-polydisperse-drag-compare-zhang2017assessment, Gasparticle-polydisperse-case-drag-gao2009drag}.
For example, different polydisperse drag models are compared to evaluate their performance in capturing the mixing and segregation of different dispersed solid flows \cite{Gasparticle-polydisperse-drag-compare-zhang2017assessment}.
Two methods are widely taken to evaluate the drag model in the polydisperse system.
Firstly each disperse phase is directly based on the drag model developed for monodisperse flow \cite{Gasparticle-polydisperse-KTGF-huilin2003hydrodynamics, Gasparticle-polydisperse-KTGF-fox-fan2008segregation}.
The second method is that the total drag of the whole multi-disperse system from either experiment or DNS simulation
is distributed to individual disperse phase according to interaction rule \cite{Gasparticle-polydisperse-drag-empirical-cello2010semi, Gasparticle-polydisperse-drag-LBM-YuA-rong2014lattice}.
The energy-minimization multiscale (EMMS) theory has been systematically developed for gas-particle flow \cite{Gasparticle-book-EMMS-li1994particle, Gasparticle-subgridmodel-EMMS-drag-yang2003cfd, Gasparticle-subgridmodel-EMMS-drag-wang2007simulation}, and been extended to the polydisperse particular flow system, with a preferred hydrodynamic performance in the EE approach \cite{Gasparticle-polydisperse-drag-EMMS-wangjunwu2019emms}.

Recently, a multiscale GKS-UGKWP method for gas-particle two-phase flow has been proposed,
which is a regime-adaptive method and can recover the EE and EL approaches in the limiting condition \cite{WP-six-gas-particle-yang2021unified, WP-gasparticle-dense-yang2022unified, WP-gasparticle-fluidized-yang2023unified}.
Unified gas-kinetic wave-particle method (UGKWP) for the particle phase is the wave-particle version of the unified gas-kinetic scheme (UGKS).
UGKS is a multiscale method, which directly models the flow physics on the scale of cell size and time step
and models the dynamics according to the cell's \text{Kn} \cite{UGKS-xu2010unified, UGKS-book-framework-xu2021cambridge}.
The method is initially developed for rarefied flow and further extended to radiation transfer, plasma, particular flow, etc \cite{UGKS-radiative-sun2015asymptotic, UGKS-plasma-liu2017unified, UGKS-gas-particle-liu2019unified}.
Different from the UGKS method with the updates of both macroscopic variables and microscopic distribution in a deterministic way,
UGKWP method updates the distribution function by analytical wave and statistical particle with a Knudsen number (\text{Kn}) dependent weights,
such as $(1-\exp(-1/\text{Kn}))$ for wave component and $\exp(-1/\text{Kn})$ for particle component.
The method increases its computational efficiency greatly in the high-speed and high-temperature flow simulation, especially close to the
equilibrium flow regime \cite{WP-first-liu2020unified, WP-second-zhu-unstructured-mesh-zhu2019unified}.
In the continuum flow regime, UGKWP gets back to the gas-kinetic scheme (GKS) \cite{GKS-2001, GKS-HLLC-compare-yang2022comparison, UGKS-book-framework-xu2021cambridge}, which is the kinetic theory based second-order NS solver. The GKS is also used to solve the gas phase
in gas-particle system. Due to the \text{Kn}-dependent wave-particle decomposition,
UGKWP is suitable for the simulation of particle flow.
In the high particle collision regime with a small \text{Kn}, no particles will be sampled in UKGWP and thus a hydrodynamic formulation will be emerged for the evolution of the solid particle phase. The whole GKS-UGKWP goes back to the EE approach.
On the contrary, when the \text{Kn} is large, such as the collisionless regime for the solid particle phase,
the evolution of the solid phase will be determined by tracking the discrete particles, and the GKS-UGKWP automatically returns to the EL approach.
At an intermediate \text{Kn}, both hydrodynamic wave and microscopic discrete particles will be updated in UGKWP for capturing the local non-equilibrium particle flow.
In this paper, for the first time the GKS-UGKWP is extended to solve polydisperse gas-particle flow with multiple disperse particle phase, where
particle's transport in the gas flow and particle collisions between same-type and different types will be incorporated into the scheme.

This paper is organized as follows. Section 2 introduces the governing equations for the particle phase and UGKWP method.
Section 3 presents the governing equations for the gas phase and GKS method.
Section 4 shows the numerical examples and the realistic engineering applications with experimental measurements.
The last section is the conclusion.

\section{UGKWP for disperse solid particle phase}
\subsection{Governing equation for disperse phase}
The evolution of disperse phase is governed by the kinetic equation,
\begin{gather}\label{disperse phase kinetic equ}
\frac{\partial f_{k}}{\partial t}
+ \nabla_\textbf{x} \cdot \left(\textbf{u}f_{k}\right)
+ \nabla_\textbf{u} \cdot \left(\textbf{a}f_{k}\right)
= \frac{g_{k}-f_{k}}{\tau_{k}}
+ \sum_{i=1,i\neq k}^{N} \frac{g_{ik}-f_{k}}{\tau_{ik}},
\end{gather}
where $f_{k}$ is the distribution function of $k-$th disperse phase, $\textbf{u}$ is the particle velocity, $\textbf{a}$ is the particle acceleration caused by the external force, $\nabla_\textbf{x}$ is the divergence operator with respect to space, $\nabla_\textbf{u}$ is the divergence operator with respect to velocity, $\tau_k$ is the relaxation time for the $k-$th disperse phase, and $g_{k}$ is the associated equilibrium distribution, which can be written as,
\begin{gather*}
g_{k}=\epsilon_k\rho_k\left(\frac{\lambda_k}{\pi}\right)^{\frac{3}{2}}e^{-\lambda_k \left[(\textbf{u}-\textbf{U}_k)^2\right]},
\end{gather*}
where $\epsilon_k$ is the volume fraction of $k-$th disperse phase, $\rho_k$ is the material density of $k-$th disperse phase, $\lambda_k$ is the variable relevant to the granular temperature $\theta_k$ with $\lambda_k = 1/(2\theta_k)$, and $\textbf{U}_k$ is the macroscopic velocity of $k-$th disperse phase.
The second term at the right hand side $(g_{ik}-f_{k})/{\tau_{ik}}$ stands for the cross-species collision, where $N$ is the number of the solid disperse phase, $\tau_{ik}$ is the auxiliary collision time, and $g_{ik}$ is the auxiliary equilibrium distribution,
\begin{gather*}
g_{ik}=\epsilon_k\rho_k\left(\frac{\lambda_k}{\pi}\right)^{\frac{3}{2}}e^{-\lambda_k \left[(\textbf{u}-\textbf{U}_{ik})^2\right]},
\end{gather*}
and note that the mass conservation for the cross-species collision can be satisfied automatically with the above $g_{ik}$.

The particle acceleration $\textbf{a}$ is determined by the external forces, such as the drag force $\textbf{D}$, the buoyancy force $\textbf{F}_b$, and gravity $m_k\textbf{G}$, etc. Particularly, $\textbf{D}$ and $\textbf{F}_b$ are inter-phase forces, standing for the force applied on the solid particles by the gas flow. The general form of drag force can be evaluated by the drag force model,
\begin{gather}\label{drag force model}
\textbf{D} = \frac{m_k}{\tau_{st}}\left(\textbf{U}_g-\textbf{u}\right).
\end{gather}
In the numerical simulation, the $\tau_{st}$ in Eq.\eqref{drag force model} will be closed by the drag model chosen for the solid phase, which will be introduced in detail later.
Besides, another interactive force considered is the buoyancy force, which can be modeled as,
\begin{gather}\label{buoyancy force model}
\textbf{F}_b = -\frac{m_k}{\rho_{k}} \nabla_\textbf{x} p_g,
\end{gather}
where $p_g$ is the pressure of the gas phase.

\subsection{UGKWP method}
In this subsection, the UGKWP for the evolution of disperse phase is introduced. Generally, the splitting operator is used to solve Eq.\eqref{disperse phase kinetic equ} through the following procedures within a numerical time step of solid phase $\Delta t_s$,
\begin{align}
\mathcal{L}_{d1} &:~~ \frac{\partial f_{k}}{\partial t}
+ \nabla_\textbf{x} \cdot \left(\textbf{u}f_{k}\right)
= \frac{g_{k}-f_{k}}{\tau_{k}},
\nonumber\\
\mathcal{L}_{d2} &:~~ \frac{\partial f_{k}}{\partial t}
= \sum_{i=1,i\neq k}^{N} \frac{g_{ik}-f_{k}}{\tau_{ik}},
\nonumber\\
\mathcal{L}_{d3} &:~~ \frac{\partial f_{k}}{\partial t}
+ \nabla_\textbf{u} \cdot \left(\textbf{a}f_{k}\right)
= 0.
\nonumber
\end{align}
For brevity, the variables updated by $\mathcal{L}_{d1}$, $\mathcal{L}_{d2}$, and $\mathcal{L}_{d3}$ are denoted as,
\begin{gather*}
\mathcal{L}_{d1}:\textbf{W}^{n}\to\textbf{W}^{*}, ~~~ \mathcal{L}_{d2}:\textbf{W}^{*}\to\textbf{W}^{**}, ~~~
\mathcal{L}_{d3}:\textbf{W}^{**}\to\textbf{W}^{n+1}.
\end{gather*}

Firstly, we focus on the part $\mathcal{L}_{d1}:\textbf{W}^{n}\to\textbf{W}^{*}$. The disperse phase kinetic equation without external force and cross-species collisions of solid particles is
\begin{gather*}
\frac{\partial f_{k}}{\partial t}
+ \nabla_\textbf{x} \cdot \left(\textbf{u}f_{k}\right)
= \frac{g_{k}-f_{k}}{\tau_{k}}.
\end{gather*}
For brevity, the subscript $k$ will be neglected in this subsection.
The integral solution of the kinetic equation can be written as,
\begin{equation}\label{particle phase integration solution}
f(\textbf{x},t,\textbf{u})=\frac{1}{\tau}\int_0^t g(\textbf{x}',t',\textbf{u} )e^{-(t-t')/\tau}\text{d}t'\\
+e^{-t/\tau}f_0(\textbf{x}-\textbf{u}t, \textbf{u}),
\end{equation}
where $\textbf{x}'=\textbf{x}+\textbf{u}(t'-t)$ is the trajectory of particle, $f_0$ is the initial distribution function at time $t=0$, and $g$ is the corresponding equilibrium state.
In UGKWP, both macroscopic conservative variables and microscopic distribution function will be updated under a finite volume framework. The cell-averaged macroscopic variables $\textbf{W}_i$ of cell $i$ are updated by the conservation law,
\begin{gather}
\textbf{W}_i^{*} = \textbf{W}_i^n - \frac{1}{\Omega_i} \sum_{S_{ij}\in \partial \Omega_i}\textbf{F}_{ij}S_{ij} + \textbf{S}_{i} \Delta t,
\end{gather}
where $\textbf{W}_i=\left(\epsilon_i\rho_i, \epsilon_i\rho_i \textbf{U}_i, \epsilon_i\rho_i E_i\right)$ are the cell-averaged macroscopic variables defined as,
\begin{gather*}
\textbf{W}_i = \frac{1}{\Omega_{i}}\int_{\Omega_{i}} \textbf{W}\left(\textbf{x}\right) \text{d}\Omega,
\end{gather*}
$\epsilon_{i}\rho_iE_i = \frac{1}{2}\epsilon_{i}\rho_i \textbf{U}_i^2 + \frac{3}{2}\epsilon_{i}\rho_i\theta_i$,
$\Omega_i$ is the volume of cell $i$, $\partial\Omega_i$ denotes the set of cell interfaces of the cell $i$, $S_{ij}$ is the area of the $j$-th interface of cell $i$, $\textbf{F}_{ij}$ denotes the fluxes for $\textbf{W}_i$ passing the interface $S_{ij}$. The flux $\textbf{F}_{ij}$ in one step $\Delta t$ can be calculated by,
\begin{align}\label{particle phase Flux equation}
\textbf{F}_{ij}=\int_{0}^{\Delta t} \int \textbf{u}\cdot\textbf{n}_{ij} f_{ij}(\textbf{x},t,\textbf{u}) \bm{\psi} \text{d}\textbf{u}\text{d}t,
\end{align}
where $\textbf{n}_{ij}$ is the unit normal vector of interface $S_{ij}$, $f_{ij}\left(t\right)$ is the distribution function on the interface $S_{ij}$, and $\bm{\psi}=(1,\textbf{u},\displaystyle \frac{1}{2}\textbf{u}^2)^T$.
Here,
$$\textbf{S}_{i} = \left[0,\textbf{0},-\frac{Q_{i,loss}}{\tau_k}\right]^T$$
stands for the lost energy due to the inelastic collision of solid particles,
\begin{gather*}
Q_{i,loss} = \left(1-e^2\right)\frac{3}{2}\epsilon_{i} \rho_{i} \theta_i,
\end{gather*}
where $e\in\left[0,1\right]$ is the restitution coefficient for the determination of the percentage of the lost energy in the inelastic collision.

Substituting the time-dependent distribution function Eq.\eqref{particle phase integration solution} into Eq.\eqref{particle phase Flux equation}, the fluxes can be rewritten as,
\begin{align*}
\textbf{F}_{ij}
&=\int_{0}^{\Delta t} \int \textbf{u}\cdot\textbf{n}_{ij} f_{ij}(\textbf{x},t,\textbf{u}) \bm{\psi} \text{d}\textbf{u}\text{d}t\\
&=\int_{0}^{\Delta t} \int\textbf{u}\cdot\textbf{n}_{ij} \left[ \frac{1}{\tau}\int_0^t g(\textbf{x}',t',\textbf{u})e^{-(t-t')/\tau}\text{d}t' \right] \bm{\psi} \text{d}\textbf{u}\text{d}t\\
&+\int_{0}^{\Delta t} \int\textbf{u}\cdot\textbf{n}_{ij} \left[ e^{-t/\tau}f_0(\textbf{x}-\textbf{u}t,\textbf{u})\right] \bm{\psi} \text{d}\textbf{u}\text{d}t\\
&\overset{def}{=}\textbf{F}^{eq}_{ij} + \textbf{F}^{fr}_{ij}.
\end{align*}

The procedure of obtaining the local equilibrium state $g_0$ at the cell interface and the construction of $g\left(t\right)$ are the same as that in GKS. For a second-order accuracy, the equilibrium state $g$ around the cell interface is written as,
\begin{gather*}
g\left(\textbf{x}',t',\textbf{u}\right)=g_0\left(\textbf{x},\textbf{u}\right)
\left(1 + \overline{\textbf{a}} \cdot \textbf{u}\left(t'-t\right) + \bar{A}t'\right),
\end{gather*}
where $\overline{\textbf{a}}=\left[\overline{a_1}, \overline{a_2}, \overline{a_3}\right]^T$, $\overline{a_i}=\frac{\partial g}{\partial x_i}/g$, $i=1,2,3$, $\overline{A}=\frac{\partial g}{\partial t}/g$, and $g_0$ is the local equilibrium on the interface.
Specifically, the coefficients of spatial derivatives $\overline{a_i}$ can be obtained from the corresponding derivatives of the macroscopic variables,
\begin{equation*}
\left\langle \overline{a_i}\right\rangle=\partial \textbf{W}_0/\partial x_i,
\end{equation*}
where $i=1,2,3$, and $\left\langle...\right\rangle$ means the moments of the Maxwellian distribution functions,
\begin{align*}
\left\langle...\right\rangle=\int \bm{\psi}\left(...\right)g\text{d}\textbf{u}.
\end{align*}
The coefficients of temporal derivative $\overline{A}$ can be determined by the compatibility condition,
\begin{equation*}
\left\langle \overline{\textbf{a}} \cdot \textbf{u}+\overline{A} \right\rangle = \textbf{0}.
\end{equation*}
Now, all the coefficients in the equilibrium state $g\left(\textbf{x}',t',\textbf{u}\right)$ have been determined, and its integration becomes,
\begin{gather}
f^{eq}(\textbf{x},t,\textbf{u}) \overset{def}{=} \frac{1}{\tau}\int_0^t g(\textbf{x}',t',\textbf{u})e^{-(t-t')/\tau}\text{d}t' \nonumber\\
= c_1 g_0\left(\textbf{x},\textbf{u}\right)
+ c_2 \overline{\textbf{a}} \cdot \textbf{u} g_0\left(\textbf{x},\textbf{u}\right)
+ c_3 A g_0\left(\textbf{x},\textbf{u}\right),
\end{gather}
with coefficients,
\begin{align*}
c_1 &= 1-e^{-t/\tau}, \\
c_2 &= \left(t+\tau\right)e^{-t/\tau}-\tau, \\
c_3 &= t-\tau+\tau e^{-t/\tau}.
\end{align*}
So, the flux from the equilibrium state $\textbf{F}^{eq}_{ij}$ is given by
\begin{gather*}
\textbf{F}^{eq}_{ij}
=\int_{0}^{\Delta t} \int \textbf{u}\cdot\textbf{n}_{ij} f_{ij}^{eq}(\textbf{x},t,\textbf{u})\bm{\psi}\text{d}\textbf{u}\text{d}t.
\end{gather*}

Besides, the flux contribution from the particle’s free transport is calculated by tracking the particles sampled from $f_0$. Therefore, the updating of the cell-averaged macroscopic variables can be written as,
\begin{gather}\label{particle phase equ_updateW_ugkp}
\textbf{W}_i^{*} = \textbf{W}_i^n - \frac{1}{\Omega_i} \sum_{S_{ij}\in \partial \Omega_i}\textbf{F}^{eq}_{ij}S_{ij}
+ \frac{\textbf{w}_{i}^{fr}}{\Omega_{i}}
+ \textbf{S}_{i} \Delta t,
\end{gather}
where $\textbf{w}^{fr}_i$ is the net free streaming flow of cell $i$, obtained by counting the sampled particle, and it stands for the flux contribution of the free streaming of particles.

The evolution of the particle distribution can be written as,
\begin{equation}
f(\textbf{x},t,\textbf{u})
=\left(1-e^{-t/\tau}\right)g^{+}(\textbf{x},t,\textbf{u})
+e^{-t/\tau}f_0(\textbf{x}-\textbf{u}t,\textbf{u}),
\end{equation}
where $g^{+}$ is named as the hydrodynamic distribution function with the analytical formulation. The initial distribution function $f_0$ has a probability of $e^{-t/\tau}$ to free transport and $(1-e^{-t/\tau})$ to collide with other particles.
The post-collision particle satisfies the distribution $g^+\left(\textbf{x},\textbf{u},t\right)$. The free transport time before the first collision with other particles is denoted as $t_c$, and then the cumulative distribution function of $t_c$ is,
\begin{gather}\label{particle phase wp cumulative distribution}
F\left(t_c < t\right) = 1 - e^{-t/ \tau},
\end{gather}
and therefore $t_c$ can be sampled as $t_c=-\tau \text{ln}\left(\eta\right)$, where $\eta$ is a random number generated from a uniform distribution $U\left(0,1\right)$. Then, the free streaming time $t_f$ for each particle is determined separately by,
\begin{gather}
t_f = min\left[-\tau\text{ln}\left(\eta\right), \Delta t\right],
\end{gather}
where $\Delta t$ is the time step. Therefore, within one time step, all particles can be divided into two groups: the collisionless particle and the collisional particle, and they are determined by the relation between time step $\Delta t$ and free streaming time $t_f$.
Specifically, if $t_f=\Delta t$, this particle is collisionless, and its trajectory is fully tracked in the whole time step.
On the contrary, if $t_f<\Delta t$, this particle is a collisional one, and its trajectory is tracked until $t_f$.
The collisional particle will be eliminated at $t_f$ in the simulation and the associated mass, momentum, and energy carried by this particle are merged into the macroscopic quantities in the relevant cell by counting its contribution through the fluxes across the cell interfaces.
More specifically, the particle trajectory in the free streaming process within time interval $t\in [0, t_f]$ is tacked by,
\begin{gather}
\textbf{x}^{*} = \textbf{x}^n + \textbf{u}^n t_f.
\end{gather}
The term $\textbf{w}_{i}^{fr}$ can be calculated by counting the particles passing through the interfaces of cell $i$,
\begin{gather}
\textbf{w}_{i}^{fr} = \sum_{k\in P\left(\partial \Omega_{i}^{+}\right)} \bm{\phi}_k - \sum_{k\in P\left(\partial \Omega_{i}^{-}\right)} \bm{\phi}_k,
\end{gather}
where $P\left(\partial \Omega_{i}^{+}\right)$ is the particle set moving into the cell $i$ within one time step, $P\left(\partial \Omega_{i}^{-}\right)$ is the particle set moving out of the cell $i$, $k$ is the particle index in the specific set,
and $\bm{\phi}_k=\left[m_{k}, m_{k}\textbf{u}_k, \frac{1}{2}m_{k}(\textbf{u}^2_k)\right]^T$ is the mass, momentum and energy carried by the particle $k$.
Therefore, $\textbf{w}_{i}^{fr}/\Omega_{i}$ is the net conservative quantities caused by the free streaming of the tracked particles. Now, all the terms in Eq.\eqref{particle phase equ_updateW_ugkp} have been determined and the macroscopic variables $\textbf{W}_i$ can be updated.

All particles have been traced up to time $t_f$. The collisionless particle with $t_f=\Delta t$ will survive at the end of the time step;
while the collisional particle with $t_f<\Delta t$ will be deleted after their first collision and it is assumed to go to the equilibrium state
in that cell. Therefore, the hydrodynamic macroscopic variables of the collisional particles in cell $i$
at the end of each time step can be directly obtained by
\begin{gather}
\textbf{W}^h_i = \textbf{W}^{*}_i - \textbf{W}^p_i,
\end{gather}
and $\textbf{W}^p_i$ are the mass, momentum, and energy of remaining collisionless particles in the cell.
Here the macroscopic variables $\textbf{W}^h_i$ account for all eliminated collisional particles,
which can be re-sampling from $\textbf{W}^h_i$ based on the Maxwellian distribution at the beginning of the next time step.
Now the updates of both macroscopic variables and the microscopic particles have been presented. The above method is the so-called unified gas-kinetic particle (UGKP) method.

The above UGKP can be further updated to UGKWP method.
In UGKP method, all particles are divided into collisionless and collisional particles in each time step.
The collisional particles are deleted after the first collision and re-sampled from $\textbf{W}^h_i$ at the beginning of the next time step.
However, only the collisionless portion of the re-samples particles can survive in the next time step,
and all re-sampled collisional ones will be deleted again.
Fortunately, the transport fluxes from these collisional particles can be evaluated analytically without using particles.
Therefore, we don't need to re-sample these collisional particles from $\textbf{W}^h_i$ at all.
According to the cumulative distribution Eq.\eqref{particle phase wp cumulative distribution}, the proportion of the collisionless particles is $e^{-\Delta t/\tau}$, and therefore in UGKWP only the collisionless particles from the hydrodynamic variables $\textbf{W}^{h}_i$ in cell $i$ will  be re-sampled with the total mass, momentum, and energy,
\begin{gather}
\textbf{W}^{hp}_i = e^{-\Delta t/\tau} \textbf{W}^{h}_i.
\end{gather}
Then, the free transport time of all these re-sampled particles will be given by $t_f=\Delta t$ in UGKWP.
The fluxes $\textbf{F}^{fr,wave}$ from these un-sampled collisional particle from $ (1- e^{-\Delta t/\tau} )\textbf{W}^{h}_i$
can be evaluated analytically \cite{WP-first-liu2020unified, WP-second-zhu-unstructured-mesh-zhu2019unified}.
Now, same as UGKP, in UGKWP the net flux $\textbf{w}_{i}^{fr,p}$ by the free streaming of the particles, which include remaining particles from the previous time step and re-sampled collisionless ones, can be calculated by
\begin{gather}
\textbf{w}_{i}^{fr,p} = \sum_{k\in P\left(\partial \Omega_{i}^{+}\right)} \bm{\phi}_k - \sum_{k\in P\left(\partial \Omega_{i}^{-}\right)} \bm{\phi}_k.
\end{gather}
So, the macroscopic flow variables in UGKWP are updated by
\begin{gather}\label{particle phase wp final update W}
\textbf{W}_i^{*} = \textbf{W}_i^n
- \frac{1}{\Omega_i} \sum_{S_{ij}\in \partial \Omega_i}\textbf{F}^{eq}_{ij}S_{ij}
- \frac{1}{\Omega_i} \sum_{S_{ij}\in \partial \Omega_i}\textbf{F}^{fr,wave}_{ij}S_{ij}
+ \frac{\textbf{w}_{i}^{fr,p}}{\Omega_{i}}
+ \textbf{S}_{i} \Delta t,
\end{gather}
where $\textbf{F}^{fr,wave}_{ij}$ is the flux function from the un-sampled collisional particles \cite{WP-first-liu2020unified, WP-second-zhu-unstructured-mesh-zhu2019unified},
\begin{align*}
\textbf{F}^{fr,wave}_{ij}
&=\textbf{F}^{fr,UGKS}_{ij}(\textbf{W}^h_i) - \textbf{F}^{fr,DVM}_{ij}(\textbf{W}^{hp}_i) \\
&=\int_{0}^{\Delta t} \int \textbf{u} \cdot \textbf{n}_{ij} \left[ e^{-t/\tau}f_0(\textbf{x}-\textbf{u}t,\textbf{u})\right] \bm{\psi} \text{d}\textbf{u}\text{d}t\\
&-e^{-\Delta t/\tau}\int_{0}^{\Delta t} \int \textbf{u} \cdot \textbf{n}_{ij} \left[g_0^h\left(\textbf{x},\textbf{u} \right) - t\textbf{u} \cdot g_\textbf{x}^h\left(\textbf{x},\textbf{u} \right) \right] \bm{\psi}\text{d}\textbf{u}\text{d}t\\
&=\int \textbf{u} \cdot \textbf{n}_{ij} \left[ \left(q_4  - \Delta t e^{-\Delta t/\tau}\right) g_0^h \left(\textbf{x},\textbf{u} \right)
+ \left(q_5 + \frac{\Delta t^2}{2}e^{-\Delta t/\tau}\right) \textbf{u} \cdot g_\textbf{x}^h\left(\textbf{x},\textbf{u} \right) \right]\bm{\psi}\text{d}\textbf{u},
\end{align*}
with,
\begin{align*}
q_4&=\tau\left(1-e^{-\Delta t/\tau}\right), \\
q_5&=\tau\Delta te^{-\Delta t/\tau} - \tau^2\left(1-e^{-\Delta t/\tau}\right).
\end{align*}

In the second part $\mathcal{L}_{d2}$, $\textbf{W}^{*}\to\textbf{W}^{**}$ models the effect of cross-species collision between solid particles
in different disperse phases.
Taking $k-$th disperse phase for example, its collision with other disperse phases can be evaluated by,
\begin{gather*}
\frac{\partial f_{k}}{\partial t}
= \sum_{i=1,i\neq k}^{N} \frac{g_{ik}-f_{k}}{\tau_{ik}}.
\end{gather*}
Obviously $\epsilon_{k}^{**}=\epsilon_k^{*}$ with the above formula of $g_{ik}$.
Taking moment $\bm{\psi}=\bm{u}$ in the Euler regime with $f_k = g_k + \mathcal{O}\left(\tau_{k}\right)$, we can obtain,
\begin{gather}
\frac{\partial \left( \epsilon_{k}\rho_k\textbf{U}_k \right)}{\partial t} = \sum_{i=1,i\neq k}^{N} \frac{\epsilon_k\rho_k\left(\textbf{U}_{ik}-\textbf{U}_k\right)}{\tau_{ik}}.
\end{gather}
In this paper, the auxiliary velocity between $i-$th and $k-$th disperse phase, $\textbf{U}_{ik}$, is assumed as,
\begin{gather}
\textbf{U}_{ik} = \frac{\epsilon_i \rho_i \textbf{U}_i + \epsilon_k \rho_k \textbf{U}_k}{\epsilon_i \rho_i + \epsilon_k \rho_k}.
\end{gather}
Now we need to determine $\tau_{ik}$, i.e., the collision time between $i-$th and $k-$th disperse phase.
Generally, the commonly employed parameter in the polydisperse particular flow is $\beta_{ik}$, which is named the so-called inter-solid drag model and has the following relationship with $\tau_{ik}$,
\begin{gather}\label{relation between tau and beta}
\frac{\epsilon_k\rho_k\left(\textbf{U}_{ik}-\textbf{U}_k\right)}{\tau_{ik}}
= \beta_{ik}\left(\textbf{U}_{i}-\textbf{U}_{k}\right).
\end{gather}
Here $\textbf{U}^{**}_k$ can be obtained by the analytical solution,
\begin{gather}
\textbf{U}_k^{**} = \left(1 - e^{- \frac{\Delta t_s}{\beta_{ik} / \epsilon_k^{*} \rho_k}}\right) \textbf{U}_i^{*} + e^{- \frac{\Delta t_s}{\beta_{ik} / \epsilon_k^{*} \rho_k}} \textbf{U}_k^{*}.
\end{gather}
The parameter $\beta_{ik}$ reflects the momentum and energy exchanges between different disperse solid phases,
which plays an important role in polydisperse solid particle flow.
Many studies have been conducted about $\beta_{ik}$ \cite{Gasparticle-polydisperse-solid-drag-syamlal1987particle, Gasparticle-polydisperse-solid-drag-mathiesen1999experimental, Gasparticle-polydisperse-KTGF-fox-fan2008segregation}.
In this paper, the inter-solid drag model proposed by Mathiesen based on KTGF will be used \cite{ Gasparticle-polydisperse-solid-drag-mathiesen1999experimental},
\begin{align*}
\beta_{ik}
& = \frac{3 p_{c,ik}}{d_{ik}} \left[ \frac{2\left(m_k^2\theta_k + m_i^2\theta_i\right)}{\pi m_0^2 \theta_k\theta_i} \right]^{1/2} \\
&  +\frac{p_{c,ik}}{|\textbf{U}_k - \textbf{U}_i|} \left[ \nabla_x \text{ln} \frac{\epsilon_{k}}{\epsilon_{i}} + 3 \nabla_x \frac{\text{ln} \left(m_i\theta_i\right)}{\text{ln} \left(m_k\theta_k\right)} +  \frac{\theta_k\theta_i}{\theta_k+\theta_i} \left( \frac{\nabla_x \theta_k}{\theta_k^2} - \frac{\nabla_x \theta_i}{\theta_i^2}\right) \right],
\end{align*}
where $p_{c,ik}$ is the collisional pressure between $i-$th and $k-$th disperse phase,
\begin{equation}\label{eq pcik}
p_{c,ik} = \frac{\pi\left(1+e_{ik}\right)d_{ik}^3g_{ik}\epsilon_i\rho_i\epsilon_k\rho_k\theta_i\theta_k\left(m_i+m_k\right)}{3\left(m_i^2\theta_i+m_k^2\theta_k\right)}\left[\frac{\left(m_i+m_k\right)^2\theta_i\theta_k}{\left(m_i^2\theta_i+m_k^2\theta_k\right)\left(\theta_i+\theta_k\right)}\right]^{3/2},
\end{equation}
with,
\begin{gather*}		
m_0=m_i+m_k,
~m_k = \frac{\pi}{6}\rho_k d_k^3,
~m_i = \frac{\pi}{6}\rho_i d_i^3, \\
~e_{ik} = \frac{e_i+e_k}{2},
~d_{ik} = \frac{d_i+d_k}{2},
g_{ik}=\frac{N}{2}\frac{\epsilon_{i} + \epsilon_{k}}{1 - \epsilon_{g}} g_0.
\end{gather*}
The influence of the cross-collision term on the granular temperature is ignored in this paper, which means $\theta_k^{**}=\theta_k^{*}$.

Finally, in the third part $\mathcal{L}_{d3}$, $\textbf{W}^{**}\to\textbf{W}^{n+1}$ accounts for the acceleration,
\begin{gather*}
\frac{\partial f_{k}}{\partial t}
+ \nabla_\textbf{u} \cdot \left(\textbf{a}f_{k}\right)
= 0,
\end{gather*}
where the acceleration of one solid particles $\textbf{a}$ can be decomposed into three parts,
\begin{gather*}
\textbf{a} = \textbf{a}_D + \textbf{a}_c + \textbf{a}_p,
\end{gather*}
where $\textbf{a}_D$ is the velocity-dependent drag force from the gas-solid interaction,
\begin{gather*}
\textbf{a}_D = \frac{\textbf{U}_g - \textbf{u}}{\tau_{st,k}},
\end{gather*}
$\textbf{a}_c$ is the velocity-independent buoyancy and gravitational forceson the solid particle,
\begin{gather*}
\textbf{a}_c = - \frac{1}{\rho_{k}} \nabla_\textbf{x} p_g + \textbf{G}.
\end{gather*}
and $\textbf{a}_p$ is the force from the collisional and frictional pressure among solid phases.
As shown later, $\textbf{a}_p$ mainly contributes in dense particle flow and has the similarity as normal stress.
It is conditionally updated in MP-PIC method \cite{Gasparticle-PIC-snider2001incompressible, Gasparticle-PIC-novel-verma2020novel, Gasparticle-PIC-snider2007three}.

Taking moment $\bm{\psi}$ on the equation of $\mathcal{L}_{d3}$, in the Euler regime with $f_k = g_k + \mathcal{O}\left(\tau_{k}\right)$,
we get
\begin{gather}\label{source term gas-solid}
\frac{\partial \textbf{W}_k}{\partial t} = \textbf{Q}_k,
\end{gather}
where
\begin{gather*}
\textbf{Q}_k=\left[\begin{array}{c}
0 \\
\frac{ \epsilon_k\rho_k \left(\textbf{U}_g - \textbf{U}_k\right)}{\tau_{st,k}}
+ \epsilon_{k}\rho_{k} \left(\textbf{a}_c + \textbf{a}_p\right) \\
\frac{\epsilon_k\rho_{k}\textbf{U}_k \cdot \left(\textbf{U}_g-\textbf{U}_k\right)}{\tau_{st,k}}
- 3\frac{\epsilon_k\rho_{k}\theta_k}{\tau_{st,k}}
+ \epsilon_{k}\rho_{k} \textbf{U}_k \cdot \left(\textbf{a}_c + \textbf{a}_p\right)
\end{array}\right].
\end{gather*}

Here, Eq.\eqref{source term gas-solid} will be updated in the following.
Firstly, the gas-solid drag between the $k-$th disperse phase and gas flow
\begin{gather*}
\left\{
\begin{array}{c}
\frac{\partial \left(\epsilon_k\rho_k \textbf{U}_k\right)}{\partial t} =
\beta_k\left(\textbf{U}_g-\textbf{U}_k\right), \\
\frac{\partial \left(\tilde{\epsilon_g} \textbf{U}_g\right)}{\partial t} = -\beta_k\left(\textbf{U}_g-\textbf{U}_k\right),
\end{array}
\right.
\end{gather*}
is discretized implicitly,
\begin{gather*}
\left\{
\begin{array}{c}
\frac{\epsilon_k^{n+1}\rho_k \textbf{U}_k^{***} - \epsilon_k^{**}\rho_k \textbf{U}_k^{**}}{\Delta t_s} =
\beta_k^{**}\left(\textbf{U}_g^{***}-\textbf{U}_k^{***}\right), \\
\frac{\tilde{\epsilon_g}^{n+1} \textbf{U}_g^{***} - \tilde{\epsilon_g}^{**} \textbf{U}_g^{**}}{\Delta t_s} =
-\beta_k^{**}\left(\textbf{U}_g^{***}-\textbf{U}_k^{***}\right),
\end{array}
\right.
\end{gather*}
where $\beta_k = \frac{\epsilon_{k}\rho_{k}}{\tau_{st,k}}$ is determined based on the drag model of $k-$th disperse phase.
Obviously we have $\epsilon_k^{n+1} = \epsilon_k^{**}$, $\tilde{\epsilon_g}^{n+1}=\tilde{\epsilon_g}^{**}$, and thus we get
\begin{gather*}
\left\{
\begin{array}{c}
\textbf{U}_k^{***} = \frac{\textbf{U}_g^{**} \Delta t_s + \textbf{U}_k^{**} r \Delta t_s + \textbf{U}_k^{**} \tau_{st,k}}{\Delta t_s + r \Delta t_s + \tau_{st,k}}, \\
\textbf{U}_g^{***} = \frac{\textbf{U}_g^{**} \Delta t_s + \textbf{U}_k^{**} r \Delta t_s + \textbf{U}_g^{**} \tau_{st,k}}{\Delta t_s + r \Delta t_s + \tau_{st,k}},
\end{array}
\right.
\end{gather*}
with $r=\frac{\epsilon_k^{**}\rho_k}{\tilde{\epsilon_g}^{**}}$. Then the particle's acceleration due to drag can be written as,
\begin{gather*}
\textbf{a}_D^{***} \overset{def}{=} \frac{\textbf{U}_g^{***} - \textbf{U}_s^{***}}{\tau_{st,k}}
= \frac{\textbf{U}_g^{**} - \textbf{U}_s^{**}}{\Delta t_s + r \Delta t_s + \tau_{st,k}},
\end{gather*}
and the acceleration without $\textbf{a}_p$ can be expressed as,
\begin{gather*}
\textbf{a}^{***}
= \textbf{a}_D^{***} + \textbf{a}_c
= \frac{\textbf{U}_g^{**} - \textbf{U}_s^{**}}{\Delta t_s + r \Delta t_s + \tau_{st,k}} + \textbf{a}_c.
\end{gather*}
for which the macroscopic variables of $k-$th solid phase are updated by
\begin{gather*}
\left\{
\begin{array}{l}
\epsilon_{k}^{n+1}\rho_k\textbf{U}_k^{***}
= \epsilon_{k}^{**}\rho_k\textbf{U}_k^{**}
+ \epsilon_{k}^{**}\rho_k \textbf{a}^{***} \Delta t_s, \\
\epsilon_{k}^{n+1}\rho_kE_k^{***} =
\epsilon_{k}^{**}\rho_kE_k^{**}
+ \left(
\epsilon_{k}^{**}\rho_k \textbf{U}_k^{**} \cdot \textbf{a}^{***}
- 3\frac{\epsilon_k^{**}\rho_{k}\theta_k^{**}}{\tau_{st,k}}
\right)\Delta t_s,
\end{array}
\right.
\end{gather*}
where $\epsilon_{k}^{n+1} \rho_k E_k^{***} = \frac{1}{2}\epsilon_{k}^{n+1}\rho_k \textbf{U}_k^2 + \frac{3}{2}\epsilon_{k}^{n+1}\rho_k\theta_k^{***}$.

As in the treatment of MP-PIC method, $\textbf{a}_p$ is updated at end as \cite{Gasparticle-PIC-snider2001incompressible, Gasparticle-PIC-snider2007three},
\begin{gather}\label{ap}
\textbf{a}_p = - \frac{1}{\epsilon_k \rho_k}\nabla_\textbf{x} \left(p_{k,c} + p_{k,f}\right),
\end{gather}
where $p_{k,c}$ and $p_{k,f}$ are the collisional pressure and frictional pressure of $k-$th disperse phase, which are determined by Eq.\eqref{pcoll} and Eq.\eqref{pfric} respectively. In this paper, the $\textbf{a}_p$ obtained by Eq.\eqref{ap} is further constrained by the following stability conditions,
\begin{gather*}
\left\{
\begin{array}{c}
\left| \frac{1}{2} \textbf{a}_p \Delta t_s^2 \right| \le k_c \Delta_{cell} , \\
\left| \textbf{U}^{***}_k\Delta t_s + \frac{1}{2} \textbf{a}_p \Delta t_s^2 \right| \le k_c \Delta_{cell},
\end{array}
\right.
\end{gather*}
where $\Delta_{cell}$ is the cell size and $k_c$ is a safety factor with a value smaller than 1, such as $0.8$ used in this paper.
Now the acceleration can be fully determined as,
\begin{gather*}
\textbf{a}^{n+1}
= \textbf{a}^{***} + \textbf{a}_p.
\end{gather*}
The macroscopic velocity of $k-$th solid phase $\textbf{U}_k^{n+1}$ is updated by
\begin{gather*}
\textbf{U}_k^{n+1} = \textbf{U}_k^{***} + \textbf{a}_p \Delta t_s,
\end{gather*}
with the granular temperature $\theta_k^{n+1}=\theta_k^{***}$.

Besides, the velocity and location of the remaining free transport particles are updated as,
\begin{align}
\textbf{u}^{n+1} &= \textbf{u}^{*} + \textbf{a}^{n+1}t_f,\\
\textbf{x}^{n+1} &= \textbf{x}^{*} + \frac{1}{2} \textbf{a}^{n+1} t_f^2.\label{displacement by acceleartion term}
\end{align}
The above procedures are used to update the disperse particle phase in one time step $\Delta t_s$.

\subsection{The \text{Kn} and flow regime of solid particle phase}
The $\text{Kn}_k$ stands for the Knudsen number of $k-$th disperse particle phase, and it is defined by the ratio of collision time $\tau_{k}$ to the characteristic time scale of macroscopic flow $t_{ref}$,
\begin{gather}\label{particle phase Kn_s}
\text{Kn}_k = \frac{\tau_k}{t_{ref}}.
\end{gather}
 The characteristic time $t_{ref}$ takes the time step of solid phase $\Delta t_s$ and $\tau_k$ is the time interval between collisions of solid particles.
 In this paper, $\tau_k$ is defined as \cite{Gasparticle-MOM-Fox-passalacqua2010fully, Gasparticle-momentmethod-Fox2013computational},
\begin{gather}\label{particle phase tau_s}
\tau_k = \frac{\sqrt{\pi}d_k}{12\epsilon_k g_0 \sqrt{\theta_k}}  ,
\end{gather}
where $d_k$, $\epsilon_k$, and $\theta_k$ are the diameter of the solid particle, volume fraction, and the granular temperature of $k-$th disperse phase. $g_0$ is the radial distribution function with the following form,
\begin{gather}\label{radial distribution g0}
g_0 = \frac{2-c}{2\left(1-c\right)^3},
\end{gather}
where $c=\epsilon_t/\epsilon_{s,max}$ is the ratio of the total solid volume fraction $\epsilon_{t}$ to the allowed maximum value $\epsilon_{s,max}$ for the polydisperse solid mixture. The flow regime of the $k-$th disperse phase is determined by $\text{Kn}_k$. Generally, for the dilute flow the collision frequency between solid particles is low, leading to a large $\text{Kn}_k$, and UGKWP will sample and track
the solid particles, keeping the non-equilibrium automatically. On the contrary, in the high concentration region,
the high collision frequency between particles makes the solid phase in the equilibrium state, and no particles will be sampled in UGKWP.
In the limit of the continuum flow regime with $e=1$, the above UGKWP method for Eq.\eqref{disperse phase kinetic equ} can recover the solution of the following hydrodynamic equations,
\begin{align}\label{Solid hydro equ}
\frac{\partial \left(\epsilon_k\rho_k\right)}{\partial t}
+ \nabla_\textbf{x} \cdot \left(\epsilon_k\rho_k \textbf{U}_k\right) & = 0,\nonumber \\
\frac{\partial \left(\epsilon_k\rho_k \textbf{U}_k\right)}{\partial t}
+ \nabla_\textbf{x} \cdot \left(\epsilon_k\rho_k \textbf{U}_k \textbf{U}_k + p_k \mathbb{I} \right)
& = \frac{\epsilon_{k}\rho_{k}\left(\textbf{U}_g - \textbf{U}_k\right)}{\tau_{st,k}} \\
& - \epsilon_{k} \nabla_\textbf{x} p_g
+ \epsilon_{k}\rho_{k} \textbf{G} +
\sum_{i=1,i\neq k}^{N} \beta_{ik}\left(\textbf{U}_{i}-\textbf{U}_{k}\right) , \nonumber \\
\frac{\partial \left(\epsilon_k\rho_k E_k\right)}{\partial t}
+ \nabla_\textbf{x} \cdot \left(\left(\epsilon_k\rho_k E_k + p_{k,k}\right) \textbf{U}_k \right)
&= \frac{\epsilon_{k}\rho_{k}\textbf{U}_k \cdot \left(\textbf{U}_g - \textbf{U}_k\right)}{\tau_{st,k}} \nonumber \\
& - 3\frac{\epsilon_k\rho_k\theta_k}{\tau_{st,k}}
- \epsilon_{k} \textbf{U}_k \cdot \nabla_\textbf{x} p_g
+ \epsilon_{k}\rho_{k} \textbf{U}_k \cdot \textbf{G}.\nonumber
\end{align}
In Eq.\eqref{Solid hydro equ}, $p_k$ is the pressure of the $k-$th disperse solid phase, and it is the sum of kinetic pressure $p_{k,k} = \epsilon_k\rho_k\theta_k$, collisional pressure $p_{k,c}$, and frictional pressure $p_{k,f}$. Lots of studies about the $p_{k,c}$ and $p_{k,f}$
have been done, especially for the dense particular flow \cite{Gasparticle-pressue-all-emms-dou2023effect}.
In this paper, the collisional pressure $p_{k,c}$ is calculate by,
\begin{equation}\label{pcoll}
p_{c,k} = \sum_{i=1,i\neq k}^{N} p_{c,ik},
\end{equation}
where $p_{c,ik}$ is the collisional pressure between the $i-$th and the $k-$th disperse phases, given in Eq.\eqref{eq pcik}. The $p_{f,k}$ accounts for the enduring inter-particle contacts and frictions of the $k-$th disperse phase, which plays important roles when the solid phase is near-packing. In this paper, the Johnson-Jackson model is employed \cite{Gasparticle-KTGF-pressure-friction-johnson1987frictional, Gasparticle-TFM-compressible-houim2016multiphase},
\begin{align}\label{pfric}
p_{f,k} = \left\{\begin{aligned}
&~~~~~~~~ 0 &  ,   &~~\epsilon_{t} \le \epsilon_{s,crit}, \\
&0.1 \epsilon_{k} \frac{\left(\epsilon_{t} - \epsilon_{s,crit}\right)^2}{\left(\epsilon_{s,max} -  \epsilon_{t}\right)^5}&  ,   &~~\epsilon_{t} > \epsilon_{s,crit}.
\end{aligned} \right.
\end{align}
Here $\epsilon_{s,crit}$ is the critical volume fraction of the whole solid phase. To avoid the solid volume fraction $\epsilon_{k}$ exceeding its maximum value $\epsilon_{s,max}$, i.e., the over-packing problem, the proposed flux limiting model near the packing condition is employed in the UGKWP method for the solid phase \cite{WP-gasparticle-dense-yang2022unified}.

\section{GKS for gas phase}
\subsection{Governing equations for gas phase}
The gas phase is regarded as the continuum flow and the governing equations are the Navier-Stokes equations with source terms reflecting the inter-phase interaction \cite{Gasparticle-book-gidaspow1994multiphase, Gasparticle-book-ishii2010thermo},
\begin{align}\label{gas phase macroscopic equ}
&\frac{\partial \left(\widetilde{\rho_g}\right)}{\partial t}
+ \nabla_x \cdot \left(\widetilde{\rho_g} \textbf{U}_g\right)= 0,\nonumber \\
&\frac{\partial \left(\widetilde{\rho_g} \textbf{U}_g\right)}{\partial t}
+ \nabla_x \cdot \left(\widetilde{\rho_g} \textbf{U}_g \textbf{U}_g + \widetilde{p_g}\mathbb{I}\right)
- \epsilon_{g} \nabla_x \cdot \left(\mu_g \bm{\sigma}\right)
=
p_g \nabla_x \epsilon_{g}
-\sum_{k=1}^{N} \frac{\epsilon_{k}\rho_{k}\left(\textbf{U}_g - \textbf{U}_k\right)}{\tau_{st}}
+ \rho_g \textbf{G}, \\
&\frac{\partial \left(\widetilde{\rho_g} E_g\right)}{\partial t}
+ \nabla_x \cdot \left(\left(\widetilde{\rho_g} E_g  + \widetilde{p_g}\right) \textbf{U}_g \right)
- \epsilon_{g} \nabla_x \cdot \left(\mu_g \bm{\sigma}\cdot\textbf{U}_g - \kappa \nabla_x T_g \right)=
\nonumber \\
& ~~~~~~~~~~~~~~~~~~~~~~~~~~~~~~~~~~~~~~~
- p_{g} \frac{\partial \epsilon_{g}}{\partial t}
- \sum_{k=1}^{N} \frac{\epsilon_{k}\rho_{k}\textbf{U}_k \cdot \left(\textbf{U}_g - \textbf{U}_k\right)}{\tau_{st}}
+ \sum_{k=1}^{N}\frac{3\epsilon_k\rho_k\theta_k}{\tau_{st}}
+ \rho_g \textbf{U}_g \cdot \textbf{G}, \nonumber
\end{align}
where $\widetilde{\rho_g}=\epsilon_{g}\rho_g$ is the apparent density of gas phase, $p_g=\rho_gRT_g$ is the pressure of gas phase and $\widetilde{p_g}=\widetilde{\rho_g}RT_g$. The strain rate tensor $\bm{\sigma}$ is
\begin{gather*}
\bm{\sigma} = \nabla_x\textbf{U}_g + \left(\nabla_x\textbf{U}_g\right)^T
- \frac{2}{3} \nabla_x \cdot \textbf{U}_g \mathbb{I},
\end{gather*}
and
\begin{gather*}
\mu_g = \tau_{g} p_g, ~~~~ \kappa = \frac{5}{2} R \tau_{g} p_g.
\end{gather*}
In particular, on the right-hand side in Eq.\eqref{gas phase macroscopic equ}, the term $p_{g} \nabla_x \epsilon_{g}$ is called ``nozzle" term, and the associated work term $- p_{g} \frac{\partial \epsilon_{g}}{\partial t}$ is called $pDV$ work term, since it is similar to the $pDV$ term in the quasi-one-dimensional gas nozzle flow equations \cite{Gasparticle-TFM-compressible-houim2016multiphase}. Unphysical pressure fluctuations might occur if the ``nozzle" term and $pDV$ term are not solved correctly. According to \cite{Toro2013book}, Eq.\eqref{gas phase macroscopic equ} can be written as the following form,
\begin{align}\label{gas phase macroscopic equ final}
&\frac{\partial \left(\rho_g\right)}{\partial t}
+ \nabla_x \cdot \left(\rho_g \textbf{U}_g\right)= C_{\epsilon_g}\rho_g,\nonumber \\
&\frac{\partial \left(\rho_g \textbf{U}_g\right)}{\partial t}
+ \nabla_x \cdot \left(\rho_g \textbf{U}_g \textbf{U}_g + p_g\mathbb{I} - \mu_g \bm{\sigma}\right)
=
C_{\epsilon_g} \rho_g \textbf{U}_g
- \sum_{k=1}^{N} \frac{\epsilon_{k}\rho_{k}\left(\textbf{U}_g - \textbf{U}_k\right)}{\epsilon_g \tau_{st}}
+ \frac{\rho_g\textbf{G}}{\epsilon_{g}}, \\
&\frac{\partial \left(\rho_g E_g\right)}{\partial t}
+ \nabla_x \cdot \left(\left(\rho_g E_g  + p_g\right) \textbf{U}_g
- \mu_g \bm{\sigma}\cdot\textbf{U}_g + \kappa \nabla_x T_g \right) = \nonumber \\
& ~~~~~~~~~~~~~~~~~~~~~~~~~
C_{\epsilon_g} \left(\rho_g E_g + p_g\right)
- \sum_{k=1}^{N}\frac{\epsilon_{k}\rho_{k}\textbf{U}_k \cdot \left(\textbf{U}_g - \textbf{U}_k\right)}{\epsilon_g \tau_{st}}
+ \sum_{k=1}^{N}\frac{3\epsilon_k\rho_k\theta_k}{\epsilon_g \tau_{st}}
+ \frac{\rho_g \textbf{U}_g \cdot \textbf{G}}{\epsilon_{g}}, \nonumber
\end{align}
where, $C_{\epsilon_g} = -\frac{1}{\epsilon_{g}}\frac{\text{d}\epsilon_{g}}{\text{d}t}$ with $\frac{\text{d}\epsilon_{g}}{\text{d}t}=\frac{\partial \epsilon_{g}}{\partial t}+\textbf{U}_g \cdot \nabla\epsilon_{g}$. The method to solve $C_{\epsilon_{g}}$ will be introduced later.

\subsection{GKS for gas evolution}
The gas flow is governed by the Navier-Stokes equations with the inter-phase interaction, and its solution will be obtained by
the corresponding gas-kinetic scheme (GKS), which is a limiting scheme of UGKWP in the continuum regime.
In general, the evolution of gas phase Eq.\eqref{gas phase macroscopic equ final} in one time step $\Delta t_g$ can be split into three parts,
\begin{align}
\mathcal{L}_{g1}&:~~
\left\{
\begin{array}{lr}
\frac{\partial \left(\rho_g\right)}{\partial t}
+ \nabla_x \cdot \left(\rho_g \textbf{U}_g\right)= 0, & \vspace{1ex}\\
\frac{\partial \left(\rho_g \textbf{U}_g\right)}{\partial t}
+ \nabla_x \cdot \left(\rho_g \textbf{U}_g \textbf{U}_g + p_g\mathbb{I} - \mu_g \bm{\sigma}\right)
= 0, & \vspace{1ex}\\
\frac{\partial \left(\rho_g E_g\right)}{\partial t}
+ \nabla_x \cdot \left(\left(\rho_g E_g  + p_g\right) \textbf{U}_g
- \mu_g \bm{\sigma}\cdot\textbf{U}_g + \kappa \nabla_x T_g \right) = 0, &
\end{array}
\right. \\
\nonumber\\
\mathcal{L}_{g2}&:~~
\left\{
\begin{array}{lr}
\frac{\partial \left(\rho_g\right)}{\partial t} = C_{\epsilon_g}\rho_g, & \vspace{1ex}\\
\frac{\partial \left(\rho_g \textbf{U}_g\right)}{\partial t} =
C_{\epsilon_g} \rho_g \textbf{U}_g, & \vspace{1ex}\\
\frac{\partial \left(\rho_g E_g\right)}{\partial t} =
C_{\epsilon_g} \left(\rho_g E_g + p_g\right). &
\end{array}
\right. \\
\nonumber\\
\mathcal{L}_{g3}&:~~
\left\{
\begin{array}{lr}
\frac{\partial \left(\rho_g\right)}{\partial t} = 0, & \vspace{1ex}\\
\frac{\partial \left(\rho_g \textbf{U}_g\right)}{\partial t} =
-\frac{\epsilon_{s}\rho_{s}\left(\textbf{U}_g - \textbf{U}_s\right)}{\epsilon_g \tau_{st}}
+ \frac{\rho_g\textbf{G}}{\epsilon_{g}}, & \vspace{1ex}\\
\frac{\partial \left(\rho_g E_g\right)}{\partial t} =
-\frac{\epsilon_{s}\rho_{s}\textbf{U}_s \cdot \left(\textbf{U}_g - \textbf{U}_s\right)}{\epsilon_g \tau_{st}}
+ \frac{3\epsilon_k\rho_k\theta_k}{\epsilon_g \tau_{st}}
+ \frac{\rho_g \textbf{U}_g \cdot \textbf{G}}{\epsilon_{g}}. &
\end{array}
\right.
\end{align}
The variables updated by $\mathcal{L}_{g1}$, $\mathcal{L}_{g2}$ and $\mathcal{L}_{g3}$ are denoted as,
\begin{gather*}
\mathcal{L}_{g1}:\textbf{W}^{n}\to\textbf{W}^{*}, ~~~ \mathcal{L}_{g2}:\textbf{W}^{*}\to\textbf{W}^{**}, ~~~ \mathcal{L}_{g3}:\textbf{W}^{**}\to\textbf{W}^{n+1}.
\end{gather*}
Firstly, the kinetic equation without nozzle and acceleration term $\mathcal{L}_{g1}$ for $\textbf{W}^{n}\to\textbf{W}^{*}$
for gas phase is modeled by,
\begin{equation}\label{gas phase kinetic equ without acce}
\frac{\partial f_{g}}{\partial t}
+ \nabla_x \cdot \left(\textbf{u}f_{g}\right)
= \frac{g_{g}-f_{g}}{\tau_{g}},
\end{equation}
where $\textbf{u}$ is the velocity, $\tau_g$ is the relaxation time for gas phase, $f_{g}$ is the distribution function of gas phase, and $g_{g}$ is the corresponding equilibrium state (Maxwellian distribution). The local equilibrium state $g_{g}$ can be written as,
\begin{gather*}
g_{g}=\rho_g\left(\frac{\lambda_g}{\pi}\right)^{\frac{K+3}{2}}e^{-\lambda_g\left[(\textbf{u}-\textbf{U}_g)^2+\bm{\xi}^2\right]},
\end{gather*}
where $\rho_g$ is the density, $\lambda_g$ is determined by gas temperature through $\lambda_g = \frac{m_g}{2k_BT_g}$, $m_g$ is the molecular mass, and $\textbf{U}_g$ is the macroscopic velocity of gas phase. Here $K$ is the internal degree of freedom with $K=(5-3\gamma)/(\gamma-1)$ for three-dimensional diatomic gas, where $\gamma=1.4$ is the specific heat ratio.
The collision term satisfies the compatibility condition
\begin{equation}\label{gas phase compatibility condition}
\int \frac{g_g-f_g}{\tau_g} \bm{\psi} \text{d}\Xi=0,
\end{equation}
where $\bm{\psi}=\left(1,\textbf{u},\displaystyle \frac{1}{2}(\textbf{u}^2+\bm{\xi}^2)\right)^T$, the internal variables $\bm{\xi}^2=\xi_1^2+...+\xi_K^2$, and $\text{d}\Xi=\text{d}\textbf{u}\text{d}\bm{\xi}$.

For brevity, the subscript $g$ will be neglected in this subsection.
For Eq.\eqref{gas phase kinetic equ without acce}, the integral solution of $f$ at the cell interface can be written as,
\begin{equation}\label{gas phase equ_integral1}
f(\textbf{x},t,\textbf{u},\bm{\xi})=\frac{1}{\tau}\int_0^t g(\textbf{x}',t',\textbf{u},\bm{\xi})e^{-(t-t')/\tau}\text{d}t'\\
+e^{-t/\tau}f_0(\textbf{x}-\textbf{u}t,\textbf{u},\bm{\xi}),
\end{equation}
where $\textbf{x}'=\textbf{x}+\textbf{u}(t'-t)$ is the trajectory of particles, $f_0$ is the initial gas distribution function at time $t=0$, and $g$ is the corresponding equilibrium state. The initial NS gas distribution function $f_0$ in Eq.\eqref{gas phase equ_integral1} can be constructed as
\begin{equation}\label{gas phase equ_f0}
f_0=f_0^l(\textbf{x},\textbf{u})(1-H(x))+f_0^r(\textbf{x},\textbf{u})H(x),
\end{equation}
where $H(x)$ is the Heaviside function, $f_0^l$ and $f_0^r$ are the
initial gas distribution functions on the left and right side of one cell interface.
More specifically, the initial gas distribution function $f_0^k$, $k=l,r$, is constructed as
\begin{equation*}
f_0^k=g^k\left(1+\textbf{a}^k \cdot \textbf{x}-\tau(\textbf{a}^k \cdot \textbf{u}+A^k)\right),
\end{equation*}
where $g^l$ and $g^r$ are the Maxwellian distribution functions on the left-hand and right-hand sides of a cell interface, which can be fully determined by the macroscopic conservative flow variables $\textbf{W}^l$ and $\textbf{W}^r$.
The coefficients $\textbf{a}^l=\left[a^l_1, a^l_2, a^l_3\right]^T$ and $\textbf{a}^r=\left[a^r_1, a^r_2, a^r_3\right]^T$ are related to the spatial derivatives in normal and tangential directions, which can be evaluated from the corresponding derivatives of the initial macroscopic variables,
\begin{equation*}
\left\langle a^l_i\right\rangle=\partial \textbf{W}^l/\partial x_i,
\left\langle a^r_i\right\rangle=\partial \textbf{W}^r/\partial x_i,
\end{equation*}
where $i=1,2,3$, and $\left\langle...\right\rangle$ means the moments of the Maxwellian distribution functions,
\begin{align*}
\left\langle...\right\rangle=\int \bm{\psi}\left(...\right)g\text{d}\Xi.
\end{align*}
Based on the Chapman-Enskog expansion, the non-equilibrium part of the distribution function satisfies,
\begin{equation*}
\left\langle \textbf{a}^l \cdot\textbf{u}+A^l\right\rangle = 0,~
\left\langle \textbf{a}^r \cdot\textbf{u}+A^r\right\rangle = 0,
\end{equation*}
and therefore the coefficients $A^l$ and $A^r$ can be fully determined. The equilibrium state $g$ around the cell interface is modeled as,
\begin{equation}\label{gas phase equ_g}
g=g_0\left(1+\overline{\textbf{a}}\cdot\textbf{x}+\bar{A}t\right),
\end{equation}
where $\overline{\textbf{a}}=\left[\overline{a}_1, \overline{a}_2, \overline{a}_3\right]^T$, $g_0$ is the local equilibrium of the cell interface. More specifically, $g$ can be determined by the compatibility condition,
\begin{align*}
\int\bm{\psi} g_{0}\text{d}\Xi=\textbf{W}_0
&=\int_{u>0}\bm{\psi} g^{l}\text{d}\Xi+\int_{u<0}\bm{\psi} g^{r}\text{d}\Xi, \nonumber \\
\int\bm{\psi} \overline{a_i} g_{0}\text{d}\Xi=\partial \textbf{W}_0/\partial x_i
&=\int_{u>0}\bm{\psi} a^l_i g^{l}\text{d}\Xi+\int_{u<0}\bm{\psi} a^r_i g^{r}\text{d}\Xi,
\end{align*}
$i=1,2,3$, and
\begin{equation*}
\left\langle \overline{\textbf{a}} \cdot \textbf{u}+\bar{A}\right\rangle = 0.
\end{equation*}
After determining all parameters in the initial gas distribution function $f_0$ and the equilibrium state $g$, substituting Eq.\eqref{gas phase equ_f0} and Eq.\eqref{gas phase equ_g} into Eq.\eqref{gas phase equ_integral1}, the time-dependent distribution function $f(\textbf{x}, t, \textbf{u},\bm{\xi})$ at a cell interface can be expressed as,
\begin{align}\label{gas phase equ_finalf}
f(\textbf{x}, t, \textbf{u},\bm{\xi})
&=c_1 g_0+ c_2 \overline{\textbf{a}}\cdot\textbf{u}g_0 +c_3 {\bar{A}} g_0\nonumber\\
&+\left[c_4 g^r +c_5 \textbf{a}^r\cdot\textbf{u} g^r + c_6 A^r g^r\right] (1-H(u)) \\
&+\left[c_4 g^l +c_5 \textbf{a}^l\cdot\textbf{u} g^l + c_6 A^l g^l\right] H(u) \nonumber.
\end{align}
with coefficients,
\begin{align*}
c_1 &= 1-e^{-t/\tau}, \\
c_2 &= \left(t+\tau\right)e^{-t/\tau}-\tau, \\
c_3 &= t-\tau+\tau e^{-t/\tau}, \\
c_4 &= e^{-t/\tau}, \\
c_5 &= -\left(t+\tau\right)e^{-t/\tau}, \\
c_6 &= -\tau e^{-t/\tau}.
\end{align*}
Then, the flux transport over a time step can be calculated,
\begin{align}
\textbf{F}_{ij} =\int_{0}^{\Delta t} \int\textbf{u}\cdot\textbf{n}_{ij} f_{ij}(\textbf{x},t,\textbf{u},\bm{\xi})\bm{\psi}\text{d}\Xi\text{d}t,
\end{align}
where $\textbf{n}_{ij}$ is the normal vector of the cell interface.
Then, the cell-averaged conservative variables of cell $i$ can be updated as follows,
\begin{gather}
\textbf{W}_i^{*} = \textbf{W}_i^n
- \frac{1}{\Omega_i} \sum_{S_{ij}\in \partial \Omega_i}\textbf{F}_{ij}S_{ij},
\end{gather}
where $\Omega_i$ is the volume of cell $i$, $\partial\Omega_i$ denotes the set of the interface of cell $i$, $S_{ij}$ is the area of $j$-th interface of cell $i$, $\textbf{F}_{ij}$ denotes the projected macroscopic fluxes in the normal direction, and $\textbf{W}_{g}=\left[\rho_g,\rho_g \textbf{U}_g, \rho_g E_g\right]^T$ are the cell-averaged conservative flow variables for the gas phase.

In the second part, $\mathcal{L}_{g2}:\textbf{W}^{*}\to\textbf{W}^{**}$ is about the nozzle term,
\begin{gather*}
\left\{
\begin{array}{l}
\rho_g^{**} = \rho_g^{*}
+ C_{\epsilon_g}^{*} \rho_g^{*} \Delta t_g, \\
\rho_g^{**}\textbf{U}_g^{**} = \rho_g^{*}\textbf{U}_g^{*}
+ C_{\epsilon_g}^{*} \rho_g^{*}\textbf{U}_g^{*}  \Delta t_g, \\
\rho_g^{**}E_g^{**} = \rho_g^{*}E_g^{*}
+ C_{\epsilon_g}^{*} \left(\rho_g^{*}E_g^{*} + p_g^{*}\right) \Delta t_g, \\
\end{array}
\right.
\end{gather*}
where
\begin{gather*}
C_{\epsilon_g}^{*}
= -\frac{1}{\epsilon_{g}^{n+1}}\left(\frac{\epsilon_{g}^{n+1} - \epsilon_{g}^{n}}{\Delta t_s}
+ \textbf{U}_g^{*} \cdot \nabla\epsilon_{g}^{n}\right),
\end{gather*}
with
\begin{gather*}
\epsilon_{g}^{n} = 1 - \sum_{k=1}^{N} \epsilon_{k}^{n}, ~~~ \epsilon_{g}^{n+1} = 1 - \sum_{k=1}^{N} \epsilon_{k}^{n+1}, ~~~ \nabla\epsilon_{g}^{n} = - \sum_{k=1}^{N} \nabla\epsilon_{k}^{n}.
\end{gather*}
It is worth noting that $\nabla\epsilon_{g}$ is the cell-averaged volume fraction gradient of the gas phase in the cell. Taking ${\partial \epsilon_{g}}/{\partial x}$ for example, it is calculated by,
\begin{equation}
\frac{\partial \epsilon_{g,i}}{\partial x} = \frac{\epsilon_{g,i+\frac{1}{2}} - \epsilon_{g,i-\frac{1}{2}}}{\Delta x},
\end{equation}
where $\epsilon_{g,i-\frac{1}{2}}$ and $\epsilon_{g,i+\frac{1}{2}}$ are volume fractions of the gas phase at the left and right interface of cell $i$, which can be obtained from the reconstructed $\epsilon_{s}$ at the interface based on $\epsilon_{s} + \epsilon_{g} = 1$.

In the third part, $\mathcal{L}_{g3}:\textbf{W}^{**}\to\textbf{W}^{n+1}$ is for the phase interaction,
\begin{gather*}
\mathcal{L}_{g3} :~~
\left\{
\begin{array}{l}
\frac{\partial \left(\rho_g\right)}{\partial t}
= 0, \\
\frac{\partial \left(\rho_g \textbf{U}_g\right)}{\partial t}
= - \sum_{k=1}^{N} \frac{\epsilon_{k}\rho_{k}\left(\textbf{U}_g - \textbf{U}_k\right)}{\epsilon_g \tau_{st}}, \\
\frac{\partial \left(\rho_g E_g\right)}{\partial t}
= - \sum_{k=1}^{N}\frac{\epsilon_{k}\rho_{k}\textbf{U}_k \cdot \left(\textbf{U}_g - \textbf{U}_k\right)}{\epsilon_g \tau_{st}}
+ \sum_{k=1}^{N}\frac{3\epsilon_{k}\rho_k\theta_k}{\epsilon_g \tau_{st}},
\end{array}
\right.
\end{gather*}
Obviously we have $\rho_g^{n+1}=\rho_g^{**}$. Then the second equation represents the momentum exchange between the gas phase with multi-disperse phases,
\begin{gather*}
\frac{\partial \left(\rho_g \textbf{U}_g\right)}{\partial t}
= - \sum_{k=1}^{N} \frac{\epsilon_{k}\rho_{k}\left(\textbf{U}_g - \textbf{U}_k\right)}{\epsilon_g \tau_{st}}
\overset{def}{=} - \frac{1}{\epsilon_g} \beta_t\left(\textbf{U}_g - \textbf{U}_t\right),
\end{gather*}
where $\beta_{t}$ and $\textbf{U}_t$ are the equivalent momentum transfer coefficient and velocity of the whole solid phase,
\begin{gather*}
\beta_t \overset{def}{=} \sum_{k=1}^{N}\beta_k = \sum_{k=1}^{N}\frac{\epsilon_{k} \rho_{k}}{\tau_{st,k}}, ~~~
\textbf{U}_t \overset{def}{=} \sum_{k=1}^{N} \frac{\beta_k \textbf{U}_k}{\beta_t}.
\end{gather*}
The calculation of $\beta_t$ and $\textbf{U}_k$ are based on the variables of $n+1$ state of the solid phase. For the above equation, the analytical solution of $\textbf{U}_g$ can be obtained,
\begin{gather*}
\textbf{U}_g^{n+1} = \textbf{U}_t^{n+1}
+ \left(\textbf{U}_g^{**} - \textbf{U}_t^{n+1}\right) e^{-\frac{\beta_t^{n+1} \Delta t_g}{\epsilon_{g}^{n+1} \rho_g^{n+1}}}.
\end{gather*}
Finally, the energy of the gas phase can be updated by,
\begin{gather*}
\rho_g^{n+1}E_g^{n+1} = \rho_g^{**}E_g^{**}
- \left[ \sum_{k=1}^{N}\frac{1}{\epsilon_g^{n+1}} \beta_k^{n+1} \textbf{U}_k^{n+1} \cdot \left(\textbf{U}_g^{n+1} - \textbf{U}_k^{n+1}\right)
-\sum_{k=1}^{N}\frac{3\epsilon_k^{n+1}\rho_k\theta_k^{n+1}}{\epsilon_g^{n+1} \tau_{st}} \right] \Delta t_g.
\end{gather*}
Now, the evolution of the gas phase in $\Delta t_g$ is finished.

\begin{figure}[htbp]
\centering
\subfigure{
	\includegraphics[height=10.0cm]{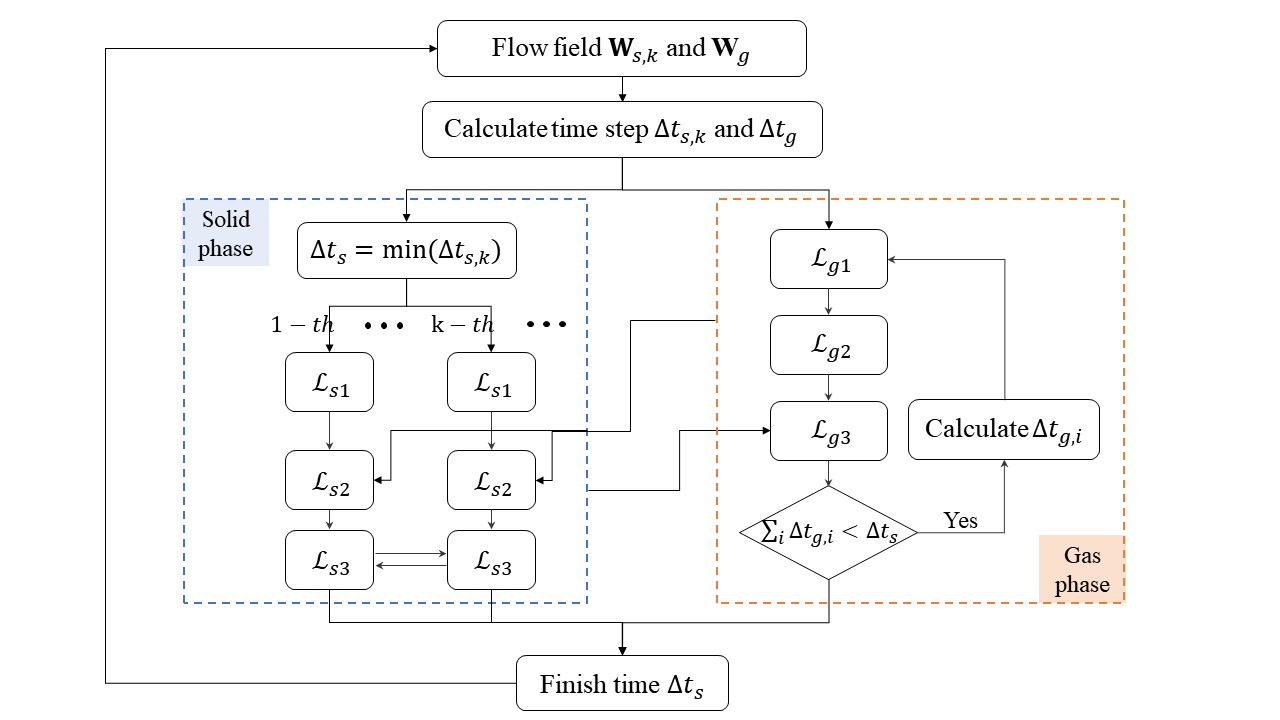}}
\caption{The flow chart of GKS-UGKWP method for polydisperse gas-particle two-phase flow.}
\label{Flow chart}
\end{figure}

In the evolution, $\Delta t_{s,k}$ and $\Delta t_g$ will be calculated based on the CFL condition; the solid phase will be updated firstly by one solid time step $\Delta t_s = \text{min}\left(\Delta t_{s,k}\right)$; then the gas phase will be updated based on the gas time step $\Delta t_g$ until $\sum_i \Delta t_{g,i} = \Delta t_s$, and the evolution of gas-particle two-phase flow in $\Delta t_s$ will be finished. The flow chart of GKS-UGKWP for polydisperse gas-particle flow is given in Figure \ref{Flow chart}.

\section{Numerical experiments}
\subsection{Circulating fluidized bed}
\subsubsection{Case description}
The first case is a circulating fluidized bed (CFB) with two disperse solid phases \cite{Gasparticle-polydisperse-case-niemi2012particle, Gasparticle-polydisperse-case-wang-niemi2015particle}.
The experiment data will be used to validate the GKS-UGKWP method.
As the previous studies \cite{ Gasparticle-polydisperse-case-wang-niemi2015particle}, the two-dimensional domain with $D\times H=0.4m\times3m$ is employed in this paper. The uniform rectangular mesh is used in the whole domain with mesh number $66\times500$, and correspondingly the cell size $\Delta_x \approx \Delta_y = 6\times10^{-3}m$. According to the experiment measurement \cite{Gasparticle-polydisperse-case-niemi2012particle}, the total inventory of solid particles is $2.85kg$, and the mass fraction, diameter and material density of each disperse phase are listed in Table \ref{particle properties of CFB}. The maximum solid volume fraction is taken as $\epsilon_{t,max} = 0.55$ in this case. Initially, the solid phase is uniformly distributed in the whole domain, and according to the mass shown in Table \ref{particle properties of CFB}, the initial solid volume fractions are $\epsilon_{1}=0.0498$ and $\epsilon_{2}=0.0140$, with the assumption of riser thickness $T=1.5cm$, which is the same as the value employed in \cite{Gasparticle-polydisperse-case-wang-niemi2015particle}. In the simulation, the solid particles are free to leave the domain at the top boundary. To simplify the simulation, the left and right boundaries are fixed walls, and thus the escaped solid particles will be replenished to the computational domain from the bottom boundary, instead of opening the right boundary adopted in
\cite{Gasparticle-polydisperse-case-wang-niemi2015particle}.
The gas with velocity $U_g=2.25m/s$ flows into the domain through the bottom boundary to fluidize the solid particles. For the left and right wall boundaries, the mixed boundary condition \cite{Gasparticle-KTGF-pressure-friction-johnson1987frictional} and no-slip wall boundary condition are used for the solid phase and gas phase, respectively.
For this case, the widely-used drag correlation proposed by Gibilaro is employed for both disperse phases \cite{Gasparticle-drag-gibilaro1985generalized, Gasparticle-drag-scaling-mckeen2003simulation}, which can be written as
\begin{equation}
\beta_{k} = \left(\frac{17.3}{Re_{s,k}}+0.336\right) \frac{\rho_g |\textbf{U}_g-\textbf{u}_k|}{d_k} \epsilon_{k}\epsilon_{g}^{-1.8},
\end{equation}
where $Re_{s,k}={\epsilon_{g}\rho_g d_k |\textbf{U}_g - \textbf{u}_k|}/{\mu_g}$ is the $Re$ of the $k$-th disperse solid phase. It is worth noting that for each disperse phase, $\tau_{st,k}$ can be obtained by the relation $\beta_{k}={\epsilon_{k} \rho_k}/{\tau_{st,k}}$.

\begin{table}[h]
	\caption{The properties of solid particles for CFB case \cite{Gasparticle-polydisperse-case-niemi2012particle}.}
	\vspace{2pt}
	\small
	\centering
	\setlength{\tabcolsep}{3.4mm}{
		\begin{tabular}{ccccc}\toprule[1pt]		
		& Total mass $\left(kg\right)$ & Mass fraction & Diameter $\left(\mu m \right)$& Material density $\left(kg/m^3\right)$ \\ \hline
		Small particle  & 2.223  & 78.0\% & 225 & 2480 \\
		Large particle & 0.627 & 22.0\% & 416 & 2480 \\
		\bottomrule[1pt]		
	\end{tabular}}
	\label{particle properties of CFB}
\end{table}

\subsubsection{Results}
\begin{figure}[htbp]
	\centering
	\subfigure{
		\includegraphics[height=6.5cm]{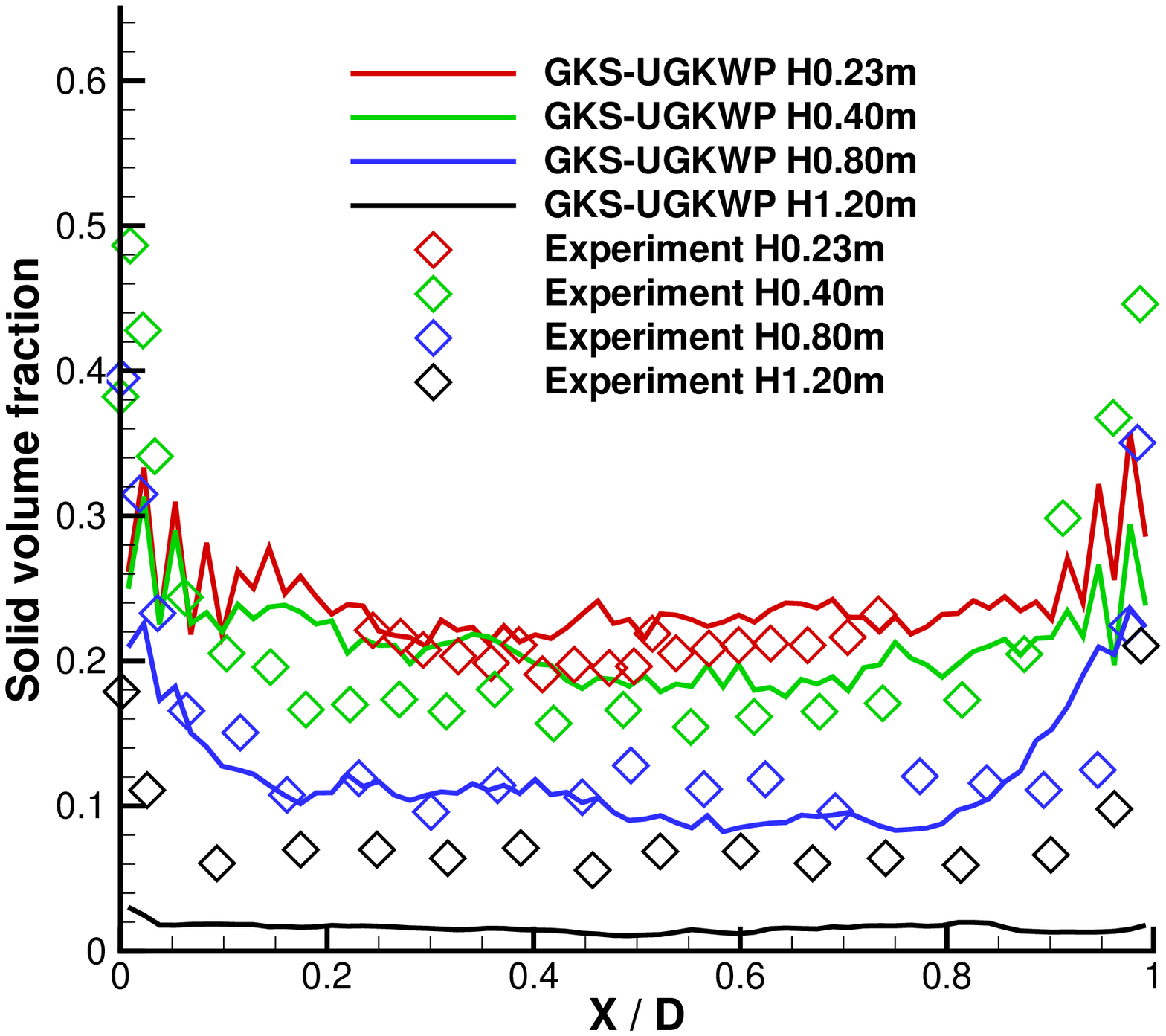}}
	\quad
	\subfigure{
		\includegraphics[height=6.5cm]{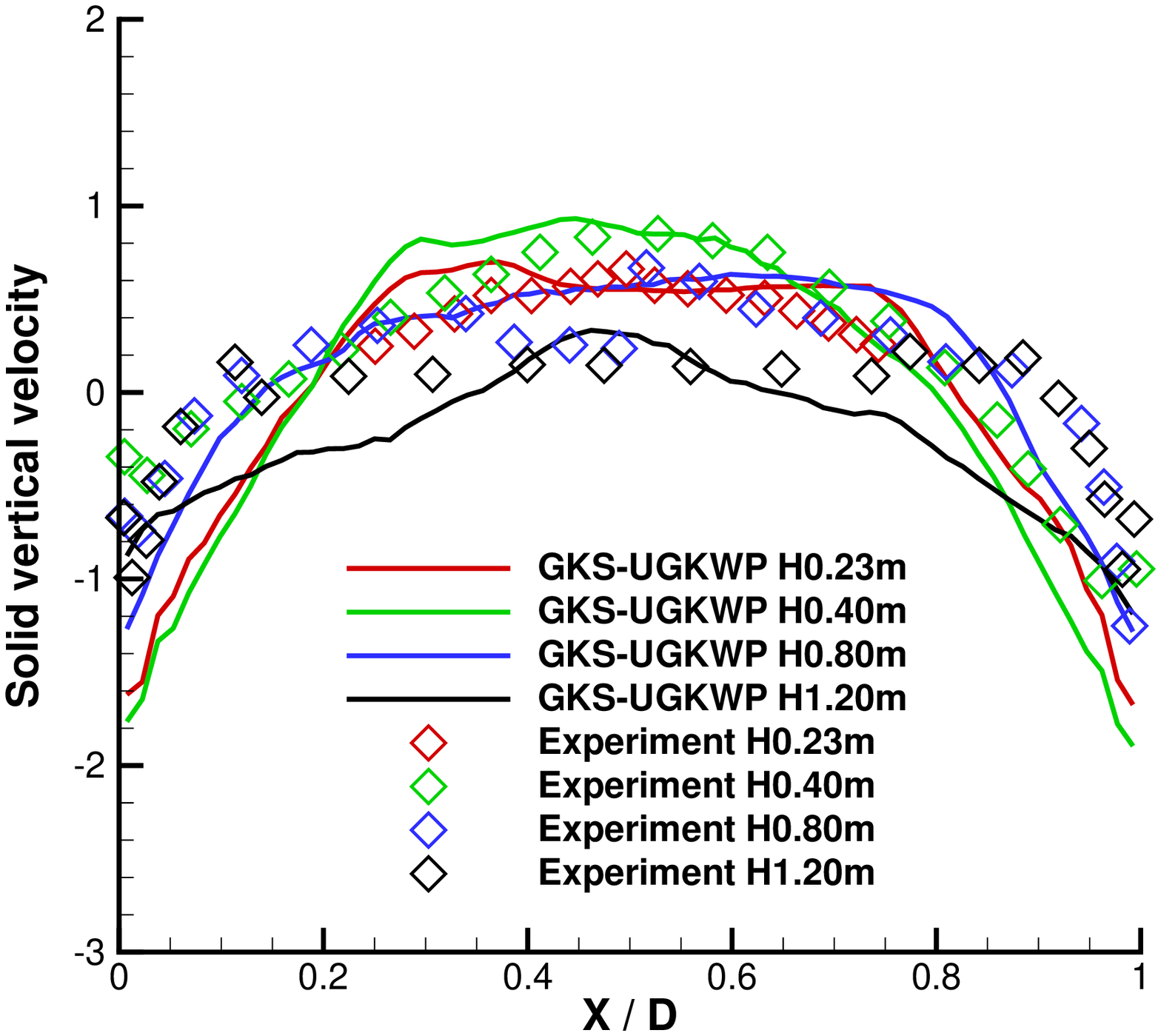}}
	\caption{The profiles of time-averaged total solid volume fraction $\epsilon_{t}$ and overall vertical velocity of solid phase $U_s$ at different riser heights by GKS-UGKWP method, and comparison with experimental measurements.}
	\label{CFB time average eps vs}
\end{figure}

\begin{figure}[htbp]
	\centering
	\subfigure[$t=6.0s$]{
		\includegraphics[height=8.0cm]{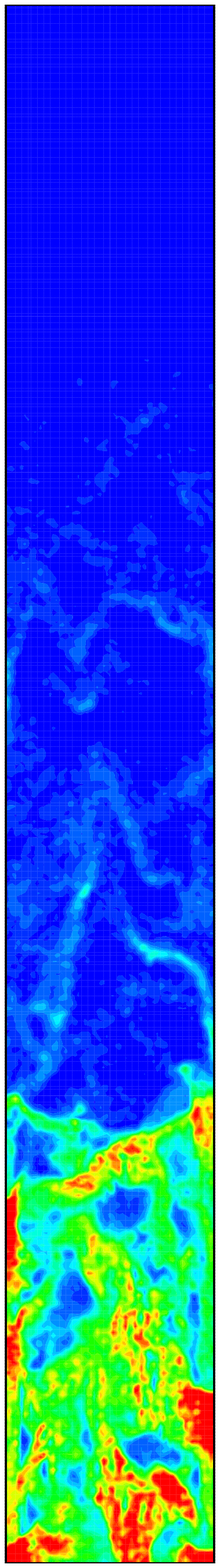}}
	\quad
	\subfigure[$t=7.0s$]{
		\includegraphics[height=8.0cm]{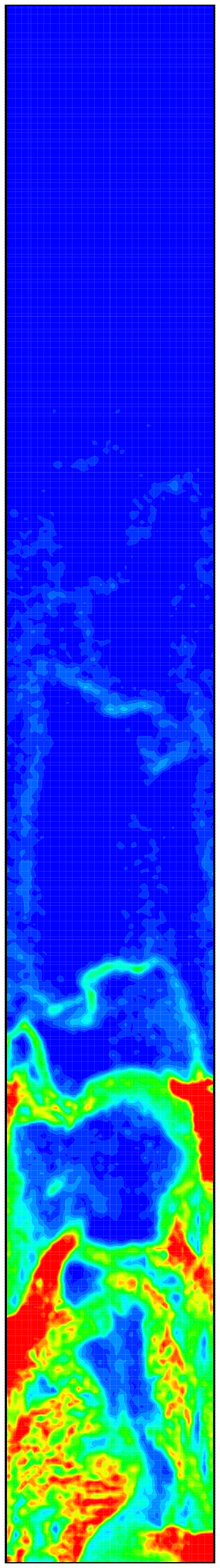}}
	\quad
	\subfigure[$t=8.0s$]{
		\includegraphics[height=8.0cm]{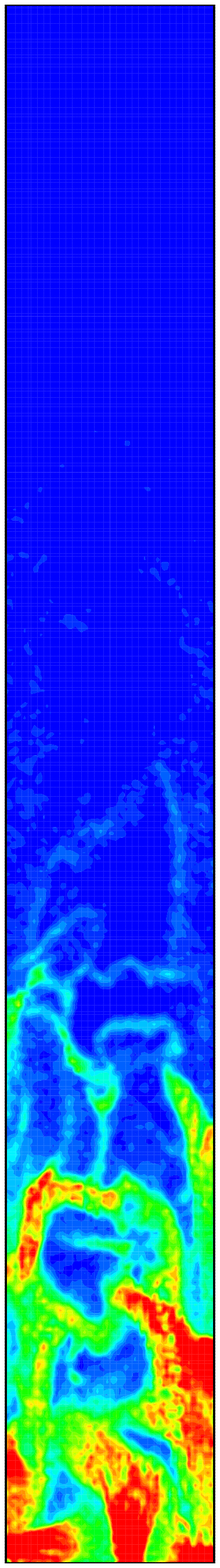}}
	\quad
	\subfigure[$t=9.0s$]{
		\includegraphics[height=8.0cm]{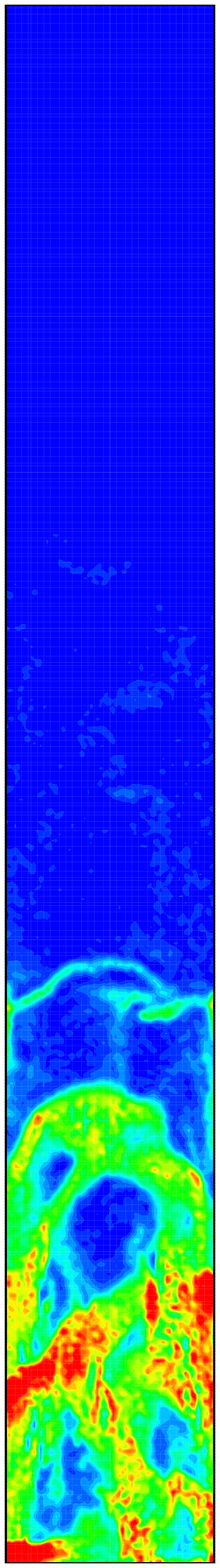}}
	\quad
	\subfigure[$t=10.0s$]{
		\includegraphics[height=8.0cm]{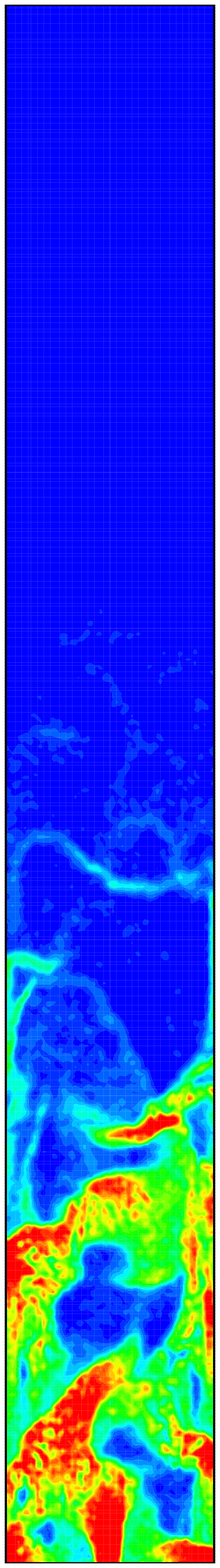}}
	\subfigure{
		\includegraphics[height=7.0cm]{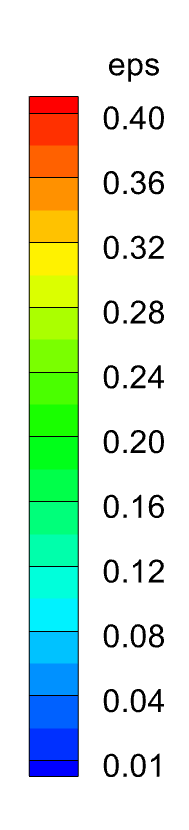}}
	\caption{The instantaneous snapshots of total solid volume fraction $\epsilon_{t}$ at time $t=6.0s, 7.0s, 8.0s, 9.0s, 10.0s$.}
	\label{CFB instantaneous eps}
\end{figure}

In this case, the simulation time is $10.0s$, and the results from $6.0s$ to $10.0s$ are used for the averaging.
Physically, to study the flow properties at different vertical positions in the riser, four gauges are set at $h=0.32m, 0.40m, 0.80m, 1.20m$ respectively in the experiment. Numerically, the total solid volume fraction $\epsilon_{t}$ and the overall vertical velocity of solid phase $U_s$ at above four heights are averaged and compared with experimental measurements in Figure \ref{CFB time average eps vs}. Note that the overall vertical velocity of the whole solid phases $U_s$ is obtained by the individual velocities weighted by solid volume fractions, $U_s=\sum_{k}\epsilon_k U_{s,k}/\sum_{k} \epsilon_k$.
Figure \ref{CFB time average eps vs} shows that the numerical prediction basically agrees with the experiment measurements.
At $h=1.20m$,  $\epsilon_t \simeq 2\%$ predicted by GKS-UGKWP is somehow lower than the experiment value $\simeq 6\%$,
which may be due to the boundary treatment, such that the escaped particles from the top boundary are replenished through the bottom boundary,
 but not from the side walls.
The snapshots of solid particles $\epsilon_{t}$ at different times are presented in Figure \ref{CFB instantaneous eps}.
In general, the solid particles prefer to accumulate at the riser's bottom and near the wall, resulting in a relatively
higher concentration in these zones. Furthermore, the instantaneous results clearly show the instantaneously coexisting
and dynamically intervening dilute/dense flow regions.
The spatially evolving solid volume fraction can be hardly captured smoothly by the hybrid EE/EL methods.
The above characteristics are also found in the studies of monodisperse CFB cases.

\begin{figure}[htbp]
	\centering
	\subfigure{
		\includegraphics[height=1.3cm]{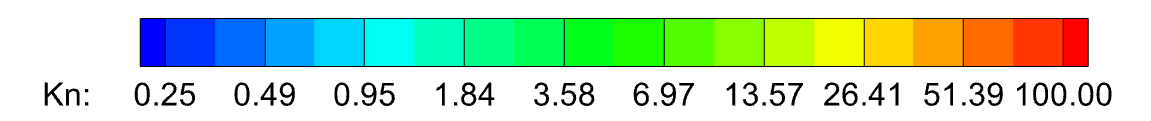}}
	\subfigure{
		\includegraphics[height=1.35cm]{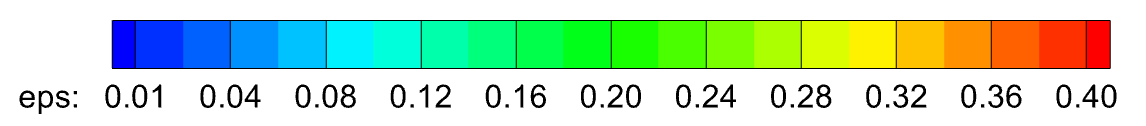}}
	\subfigure{
		\includegraphics[height=1.25cm]{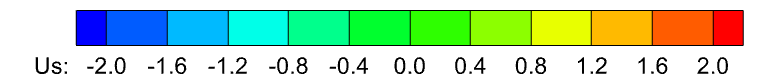}}\\
	\subfigure{
		\includegraphics[height=8.0cm]{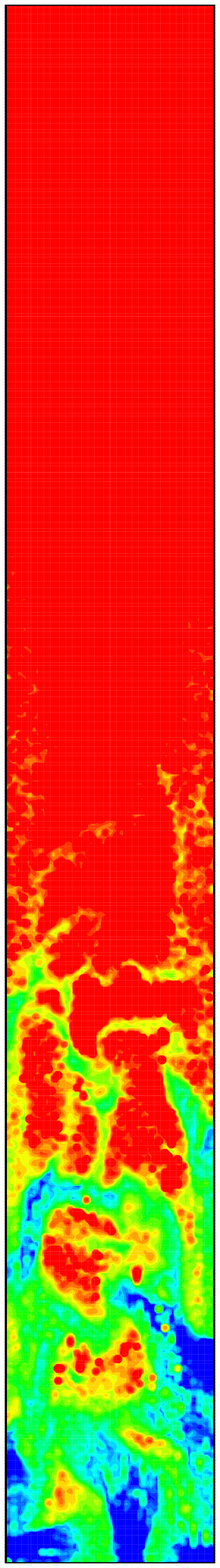}}
	\hspace{-8.7mm}
	\subfigure{
		\includegraphics[height=8.0cm]{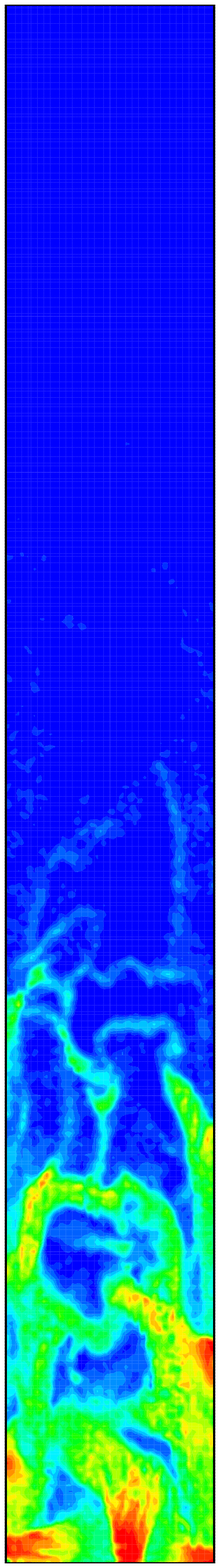}}
	\hspace{-8.7mm}
	\subfigure{
		\includegraphics[height=8.0cm]{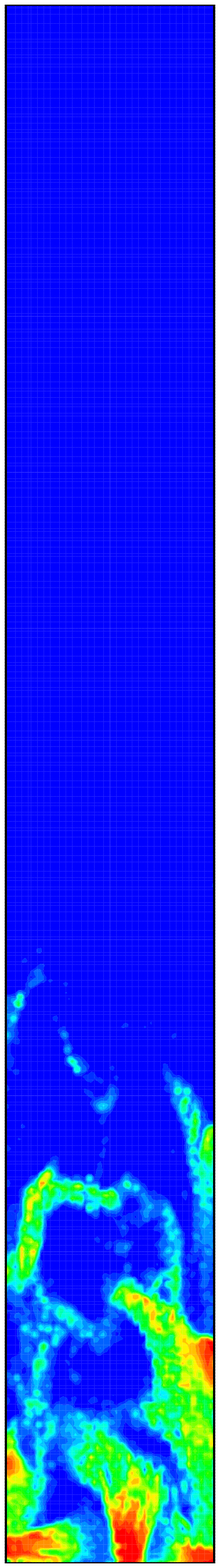}}
	\hspace{-4.7mm}
	\subfigure{
		\includegraphics[height=8.0cm]{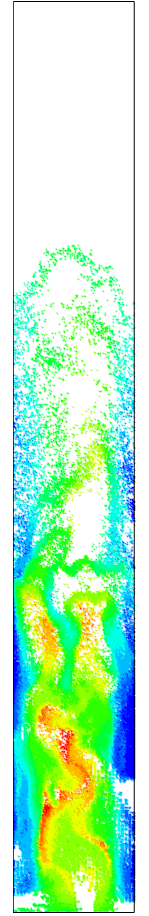}}
	\hspace{-4.7mm}
	\subfigure{
		\includegraphics[height=8.0cm]{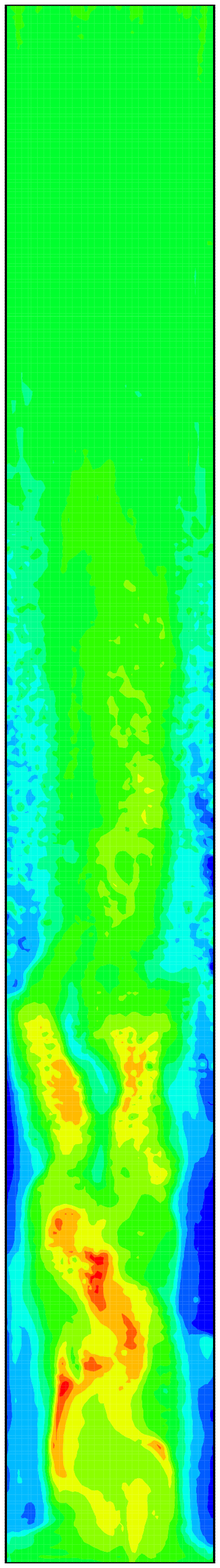}}
	\hspace{-5mm}		
	\subfigure{
		\includegraphics[height=8.0cm]{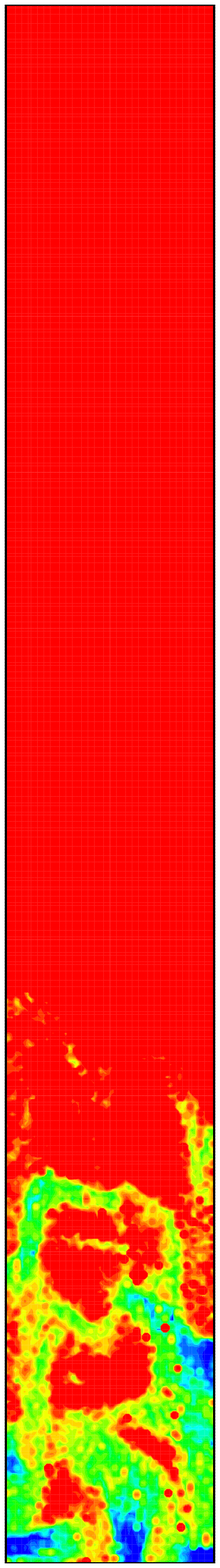}}
	\hspace{-8.7mm}
	\subfigure{
		\includegraphics[height=8.0cm]{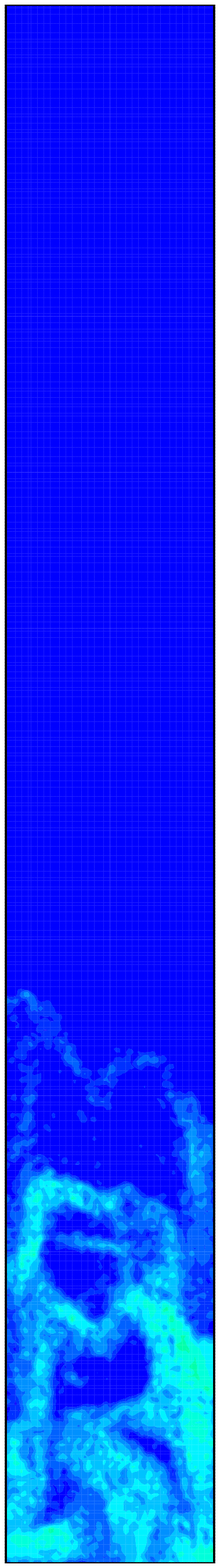}}
	\hspace{-8.7mm}
	\subfigure{
		\includegraphics[height=8.0cm]{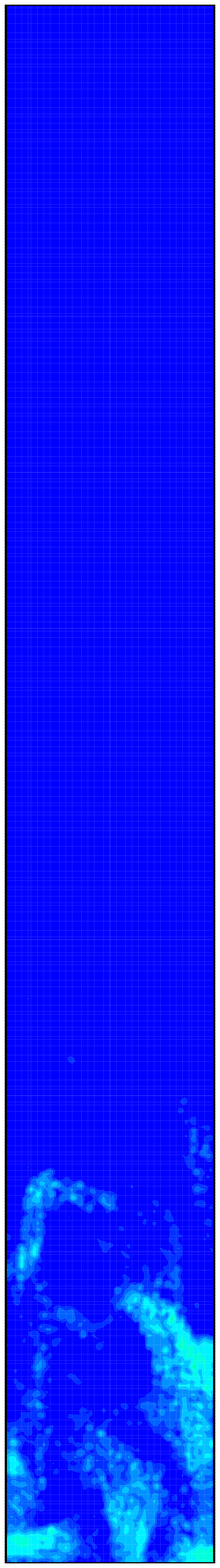}}
	\hspace{-4.7mm}
	\subfigure{
		\includegraphics[height=8.0cm]{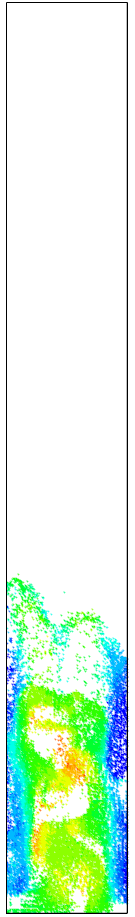}}
	\hspace{-4.7mm}
	\subfigure{
		\includegraphics[height=8.0cm]{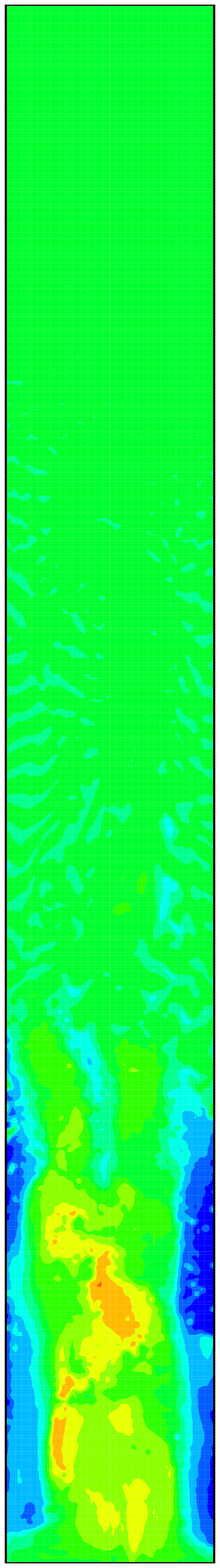}}
	\begin{center}
		\vspace{-2pt}		
		\footnotesize \quad $\text{Kn}_1$ \quad \quad~ $\epsilon_{1}$ \quad \quad~ $\epsilon_{1}^{wave}$ \quad \quad $P_1$ \qquad~ $U_{s,1}$ \quad \quad \quad~ $\text{Kn}_2$ \quad \quad~ $\epsilon_{2}$ \quad \quad~ $\epsilon_{2}^{wave}$ \quad \quad $P_2$ \qquad~ $U_{s,2}$ \quad
	\end{center}	
	\caption{The instantaneous snapshots of \text{Kn}, solid volume fraction $\epsilon$, solid volume fraction by wave in UGKWP $\epsilon^{wave}$, the set of sampled particles in UGKWP $P$, and the vertical velocity of solid phase $U_s$ at $t=8.0s$. The figures with subscript 1 are the results of $1$st solid phase (small particle), and the figures with subscript 2 are the results of $2$nd solid phase (large particle).
The $\text{Kn}_k$ is colored by the Kn-legend. The solid volume fraction $\epsilon_k$ and the corresponding wave component $\epsilon^{wave}_k$ are colored by the eps-legend. The discrete particles in particle set $P_k$ and the vertical velocity of solid phase $U_{s,k}$ are colored by the \text{Us}-legend, with $k=1,2$. The legend of \text{Kn} is in the exponential distribution. Note that the sum of $\epsilon_{1}$ and $\epsilon_{2}$ is exactly equal to $\epsilon_{t}$ at $t=8.0s$ shown in Figure \ref{CFB instantaneous eps}.}
	\label{CFB t8}
\end{figure}

For each disperse solid phase, the $\text{Kn}_k$ defined by $\text{Kn}_k=\frac{\tau_k}{\Delta t_s}$ with the local collision time  $\tau_k$ of
the $k$-th disperse phase, is presented in Figure \ref{CFB t8}.
Distributed by $\text{Kn}_k$, the wave component, contour of $\epsilon^{wave}_k$, and the particle component, the set of sampled particles colored by its vertical velocity $P_k$, are also shown in Figure \ref{CFB t8}.
Note that, the sum of the wave $\epsilon_{k}^{wave}$ and the solid particle $P_k$ components
is equal to $\epsilon_{k}$ shown in Figure \ref{CFB t8}.
The vertical velocity of each solid phase $U_{s,k}$ are also given in Figure \ref{CFB t8}.
The spatial distribution of $\epsilon$ and $U_s$ of two particle phases are distinguishable, indicating the necessity of the polydiserse method.
For both solid phases, \text{Kn} is generally smaller in the near-bottom and near-wall zones of the riser
due to the accumulation and collisions of particles in these regions.
The two disperse solid particle phases adjust their  weights to the wave and particle components in UGKWP
according to their respective \text{Kn}.
One obvious advantage of the GKS-UGKWP for polydisperse flow is that each disperse phase can take a most-efficient way for its decomposition into wave and particle.

\subsection{Turbulent fluidized bed}
\subsubsection{Case description}
The dense turbulent fluidized bed (TFB) was applied in the petroleum refining industry and was studied experimentally and numerically  \cite{Gasparticle-polydisperse-case-drag-gao2009drag, Gasparticle-polydisperse-case-gao2009hydrodynamics}.
In this problem, two kinds of particles, such as the FCC catalyst (fine) and millet (coarse), are involved with
detailed properties given in Table \ref{particle properties of TFB}.
Table \ref{particle properties of TFB} shows that the densities of two types of particles are very close,
while the particles sizes are much different. This flow condition brings challenge to the numerical methods for the gas-particle system
with single solid phase alone, where the tracking of multiple solid phases in GKS-UGKWP seems suitable for this problem.
The case of initial bed height $H_0 = 1.155m$ and gas velocity $U_g=0.53m/s$ is studied in this paper.
The computational domain is $D\times H=0.5m \times4m$ covered by uniform rectangular mesh $40\times300$.
The maximum solid volume fraction is taken as $\epsilon_{s,max}=0.65$ in this study.
At the beginning of the simulation, all solid particles are uniformly distributed in the whole riser with initial volume fraction $\epsilon_{1}=0.118$ and $\epsilon_{2}=0.070$ for the small and large particle phases, respectively.
The solid particles escaping from the top boundary will be recirculated back to the computational domain through the bottom boundary.
For the gas phase, the standard atmospheric condition is employed at the top boundary, and the gas blows into the riser through the bottom boundary with the velocity $U_g$ and a pressure difference of $\Delta p = \epsilon_{s,max} (\rho_s^* -\rho_g) GH_0$  from the top boundary, where $\rho_s^*=1463.3kg/m^3$ is the density of the solid particles weighted by their initial volume fractions.
Same as the above CFB case, the mixed boundary condition and no-slip wall boundary condition are employed on the side walls
for the particle phase and gas phase respectively.

\begin{table}[h]
	\caption{The properties of fine and coarse particles in TFB \cite{Gasparticle-polydisperse-case-gao2009hydrodynamics}.}
	\vspace{2pt}
	\small
	\centering
	\setlength{\tabcolsep}{3.4mm}{
		\begin{tabular}{ccccc}\toprule[1pt]		
			Solid phase & Mass fraction & Diameter $\left(\mu m \right)$& Material density $\left(kg/m^3\right)$ & Geldart group \\ \hline
		 	FCC catalyst & 64.2\%  & 60 & 1500 & Geldart A \\
			Millet & 35.8\% & 930 & 1402 & Geldart D \\
			\bottomrule[1pt]		
	\end{tabular}}
	\label{particle properties of TFB}
\end{table}

In GKS-UGKWP, each solid phase can choose the most accurate and suitable drag model for the polydisperse system.
As shown in Table \ref{particle properties of TFB}, the fine and coarse particles are the Geldart A and D group respectively,
and different drag models are employed to evaluate the gas-solid interaction in the FCC catalyst and millet particle phase.
Different drag models and their modifications are studied and compared \cite{Gasparticle-polydisperse-case-gao2009hydrodynamics}.
More specifically, for the coarse particle, the Gidaspow model is used \cite{Gasparticle-book-gidaspow1994multiphase},
\begin{equation}\label{drag gidaspow}
\beta_{k}=
\left\{\begin{aligned}
&150\frac{\epsilon_{k} \left(1-\epsilon_{g}\right) \mu_g}{\epsilon_{g} d_k^2} + 1.75\frac{\epsilon_{k}\rho_g |\textbf{U}_g - \textbf{u}_k|}{d_k},  & \epsilon_{g} \le 0.8,\\
&\frac{3}{4}C_d\left(Re_{s,k}\right) \frac{\epsilon_{k}\epsilon_{g}\rho_g}{d_k}|\textbf{U}_g-\textbf{u}_k| \epsilon_{g}^{-2.65},   & \epsilon_{g} > 0.8,
\end{aligned}\right.
\end{equation}
while for the fine FCC catalyst particle, the four-zone drag model is employed,
\begin{equation}\label{drag gaojs}
\beta_{k}=
\left\{\begin{aligned}
&150\frac{\epsilon_{k} \left(1-\epsilon_{g}\right) \mu_g}{\epsilon_{g} \left(d_k^*\right)^2} + 1.75\frac{\epsilon_{k}\rho_g |\textbf{U}_g - \textbf{u}_k|}{d_k^*},  & 0 \le \epsilon_{g} \le 0.8,\\
&\frac{5}{72}C_d\left(Re_{s,k}^*\right)\frac{\epsilon_{k}\epsilon_{g}\rho_g}{d_k^*\left(1 - \epsilon_{g}\right)^{0.293} }|\textbf{U}_g-\textbf{u}_k|,  & 0.8 < \epsilon_{g} \le 0.933, \\
&\frac{3}{4}C_d\left(Re_{s,k}\right) \frac{\epsilon_{k}\epsilon_{g}\rho_g}{d_k}|\textbf{U}_g-\textbf{u}_k| \epsilon_{g}^{-2.65},   & 0.933 < \epsilon_{g} \le 0.990, \\
&\frac{3}{4}C_d\left(Re_{s,k}\right) \frac{\epsilon_{k}\rho_g}{d_k}|\textbf{U}_g-\textbf{u}_k|,   & 0.990 < \epsilon_{g} \le 1.0,
\end{aligned}\right.
\end{equation}
where $d_k$ is the diameter of solid particle, and $d_k^*$ in Eq.\eqref{drag gaojs} is the effective diameter of the FCC catalyst particle \cite{Gasparticle-polydisperse-case-gao2009hydrodynamics, Gasparticle-polydisperse-case-drag-gao2009drag}.
With the consideration of particle clusters, $d_k$ is taken as $300\mu m$ for better agreement with experimental measurement \cite{Gasparticle-polydisperse-case-gao2009hydrodynamics}, which is also employed here.
Besides, in Eq.\eqref{drag gidaspow} and Eq.\eqref{drag gaojs}, the $C_d$ and $Re_{s,k}$ are defined as below,
\begin{gather*}
C_d \left(Re_k\right) =
\left\{\begin{aligned}
&\frac{24}{Re_k}\left(1 + 0.15Re_k^{0.687}\right),  & Re_k \le 1000,\\
&0.44,  & Re_k > 1000,
\end{aligned}\right.
\end{gather*}
and
\begin{gather*}
Re_{s,k} = \frac{\epsilon_{g}\rho_g d_k |\textbf{U}_g - \textbf{u}_k|}{\mu_g}, ~~~
Re_{s,k}^{*} = \frac{\epsilon_{g}\rho_g d_k^* |\textbf{U}_g - \textbf{u}_k|}{\mu_g}.
\end{gather*}

\subsubsection{Results}
\begin{figure}[htbp]
	\centering
	\subfigure{
		\includegraphics[height=6.5cm]{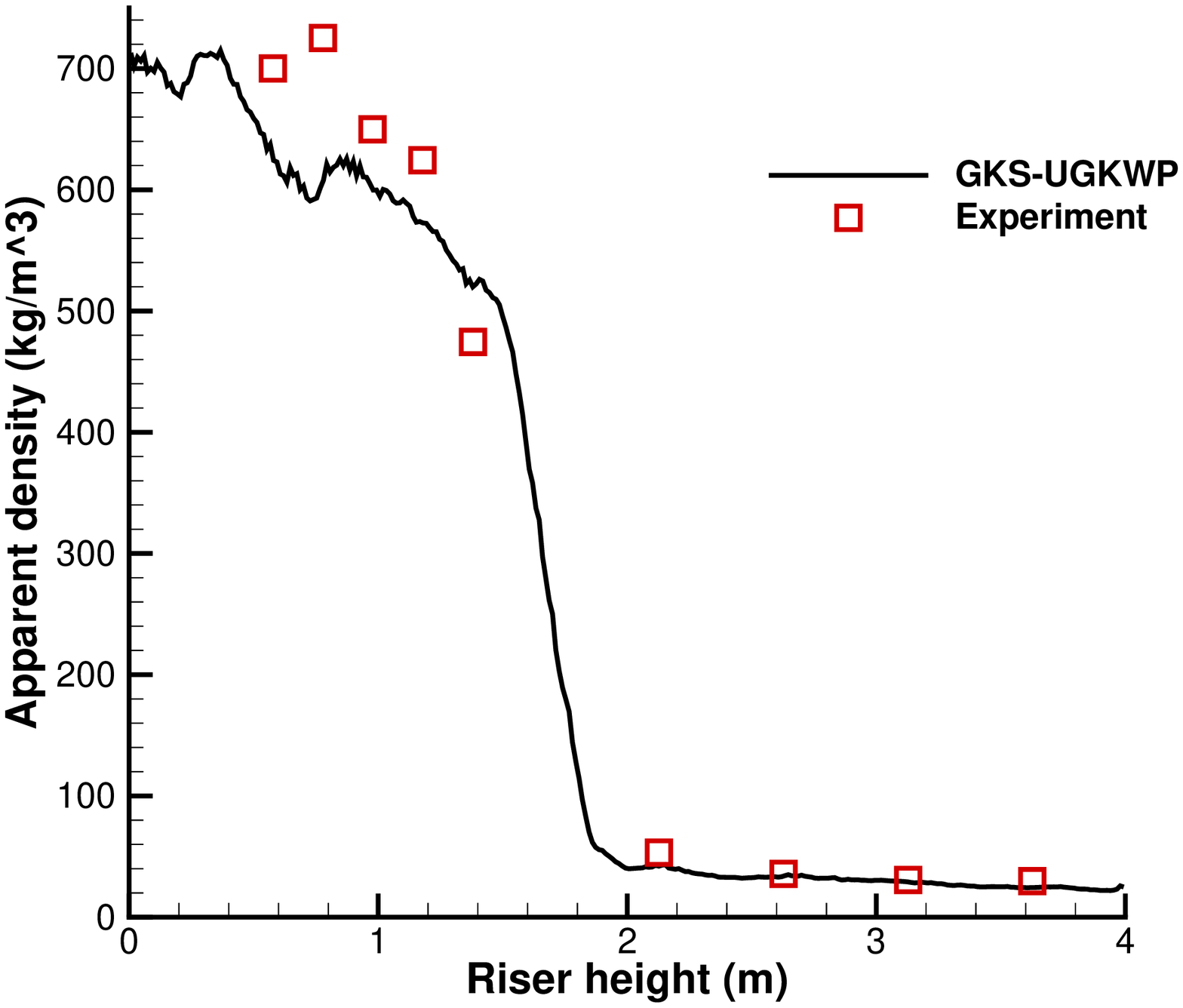}}
	\quad
	\subfigure{
		\includegraphics[height=6.5cm]{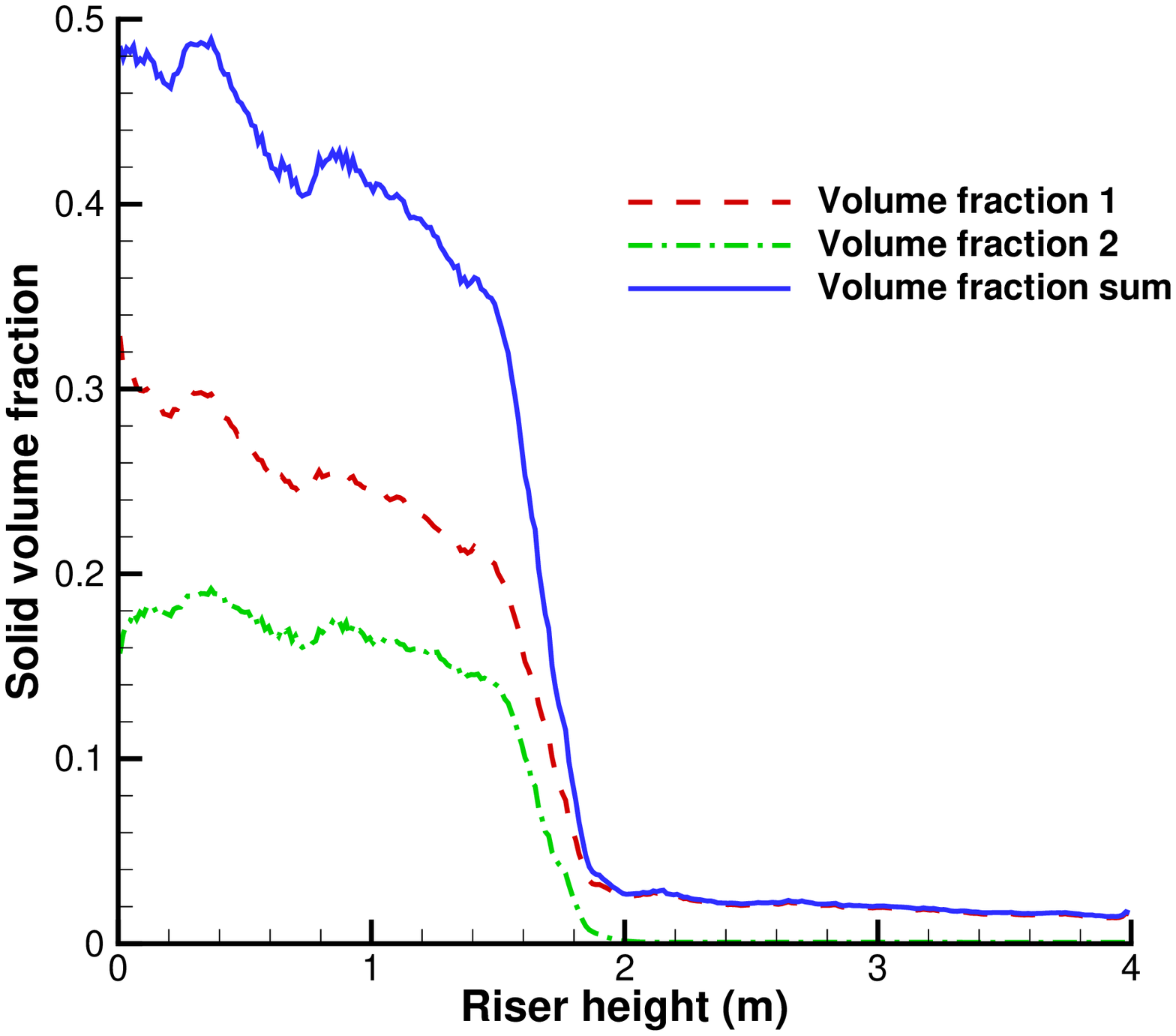}}
	\caption{Left: the profiles of time-averaged apparent density of whole solid phase $\sum_{k}\epsilon_{k}\rho_{k}$ along the riser height by GKS-UGKWP method and comparison with experimental measurements. Right: the profiles of time-averaged solid volume fraction of each disperse phase $\epsilon_{k}, k=1,2,$ and their sum $\epsilon_{t}$.}
	\label{TFB time average eps rhos}
\end{figure}

The time-averaged results from $10.0s$ to $15.0s$ are shown in Figure \ref{TFB time average eps rhos}.
The solid phase in the riser shows higher concentration at the bottom zone and lower density in the top zone,
and the transition region is very small, approximately $1.7m \sim 2.0m$.
Overall, the predicted apparent density of the solid phase agrees well with the experiment measurements.
Besides, Figure \ref{TFB time average eps rhos} also presents profiles of $\epsilon$ for two solid phases and
shows a similar trend along the riser height.

\begin{figure}[htbp]
	\centering
	\subfigure[$t=10.0s$]{
		\includegraphics[height=8.0cm]{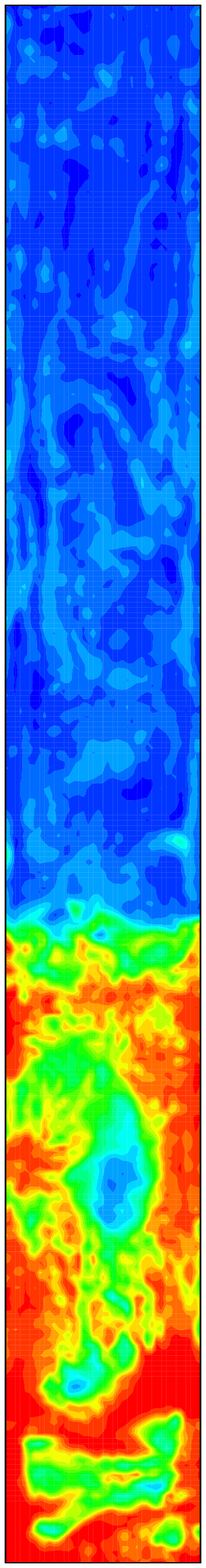}}
	\subfigure[$t=11.0s$]{
		\includegraphics[height=8.0cm]{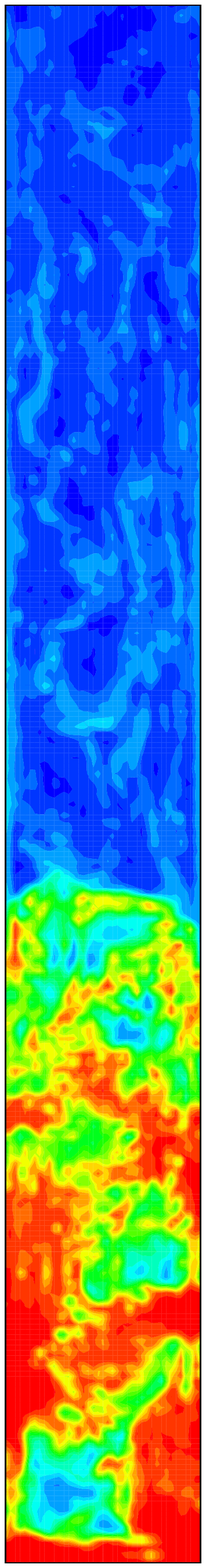}}
	\subfigure[$t=12.0s$]{
		\includegraphics[height=8.0cm]{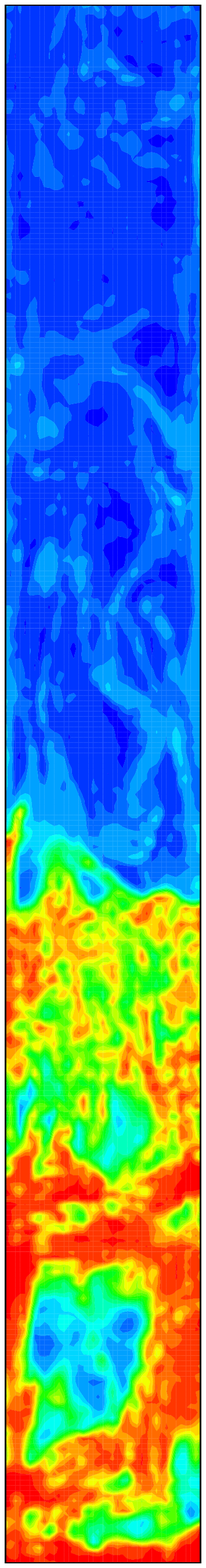}}
	\subfigure[$t=13.0s$]{
		\includegraphics[height=8.0cm]{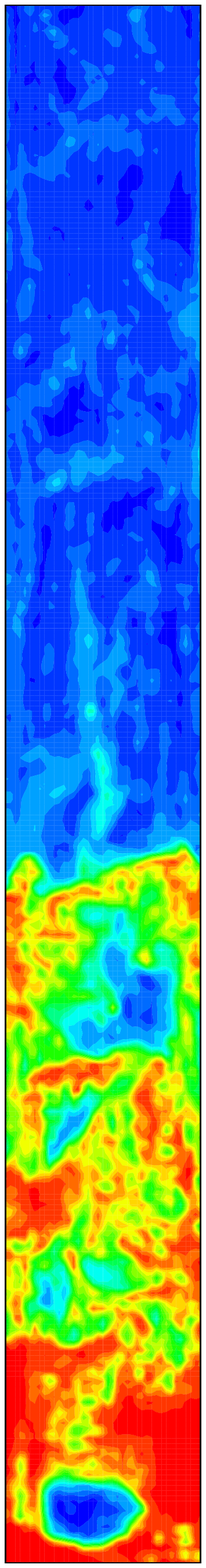}}
	\subfigure[$t=14.0s$]{
		\includegraphics[height=8.0cm]{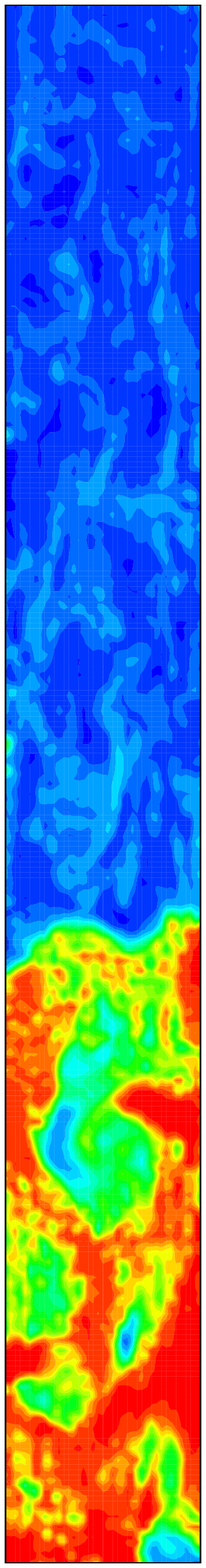}}
	\subfigure[$t=15.0s$]{
		\includegraphics[height=8.0cm]{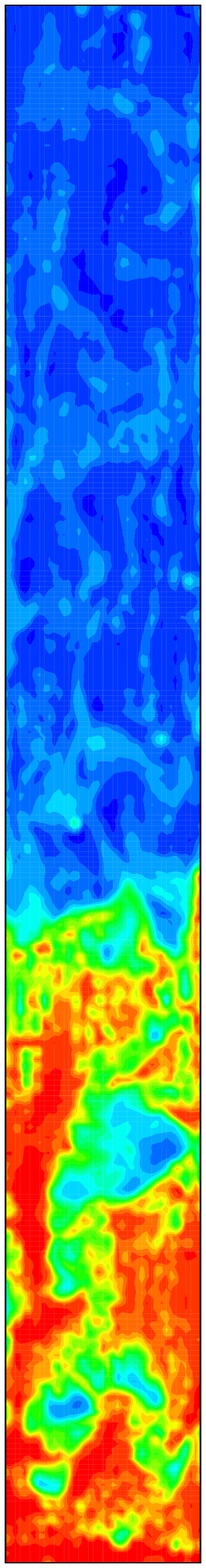}}
	\subfigure{
		\includegraphics[height=7.0cm]{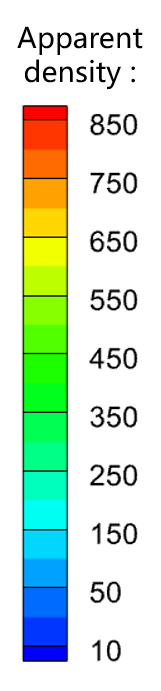}}
	\caption{The instantaneous snapshots of apparent density of the whole solid phase $\sum_{k}\epsilon_{k}\rho_{k}$ at $t=10.0 \sim 15.0s$.}
	\label{TFB instantaneous rhos}
\end{figure}

The instantaneous snapshots of the solid phase density, $\sum_{k}\epsilon_{k}\rho_{k}$, from $10.0s\sim15.0s$ are given in Figure \ref{TFB instantaneous rhos},
which indicates a typical feature of TFB. The coexistence pattern of bottom dense/middle transition/up dilute regions is well captured.
All solid particles in the top dilute region are FCC catalyst (Geldart A) type.
Figure \ref{TFB instantaneous rhos} shows the particle cluster phenomenon, which have difficulty in the drag modeling in these regions.
In the bottom dense region, the gas bubble with variable solid particles inside shows complex pattern and its tangling with the dense solid phase.

\begin{figure}[htbp]
	\centering
	\subfigure{
		\includegraphics[height=1.3cm]{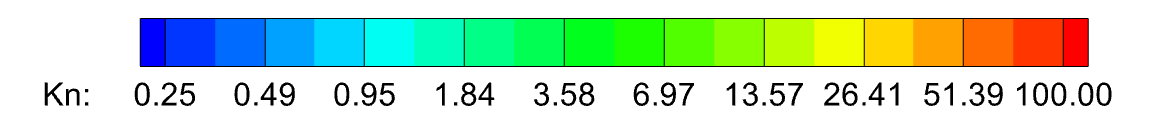}}	
	\subfigure{
		\includegraphics[height=1.15cm]{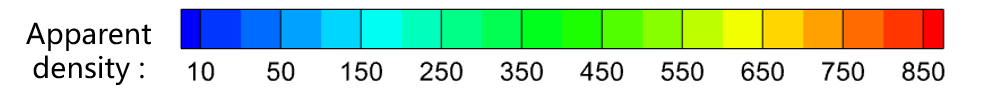}}
	\subfigure{
		\includegraphics[height=1.15cm]{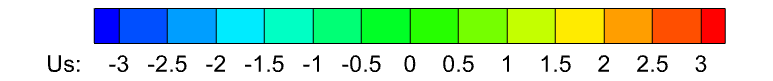}}\\
	\subfigure{
		\includegraphics[height=8.0cm]{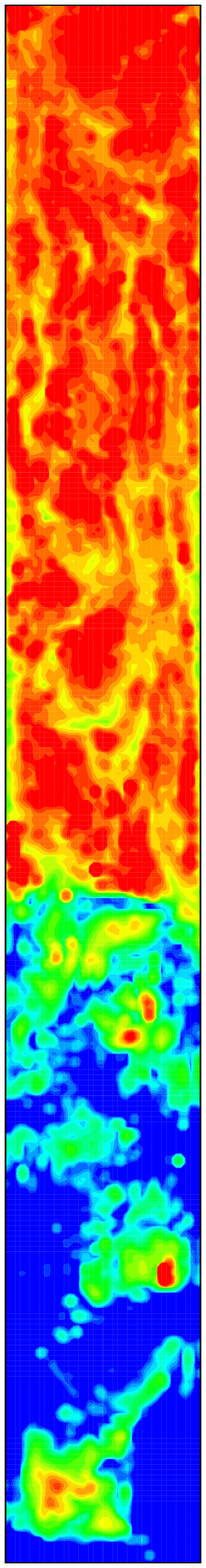}}
	\hspace{-5mm}
	\subfigure{
		\includegraphics[height=8.0cm]{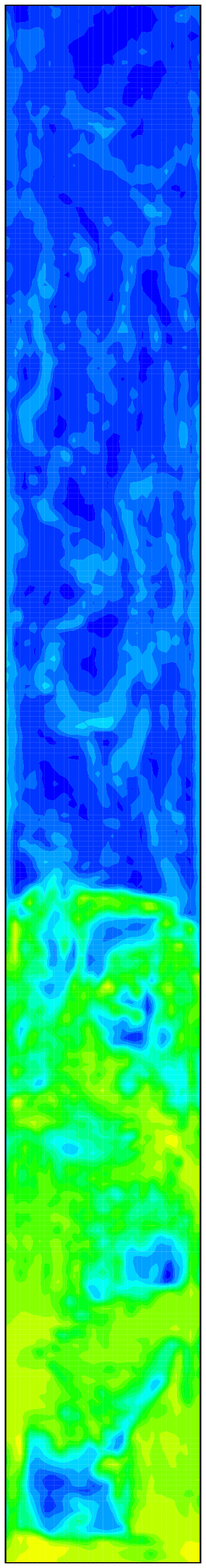}}
	\hspace{-5mm}
	\subfigure{
		\includegraphics[height=8.0cm]{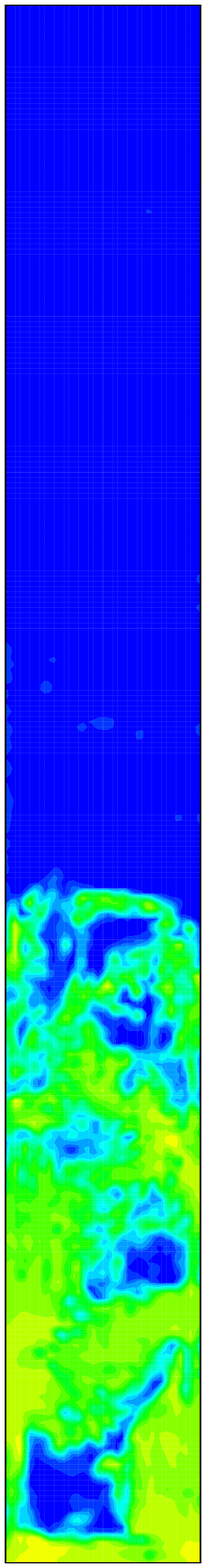}}
	\hspace{-1.5mm}
	\subfigure{
		\includegraphics[height=8.0cm]{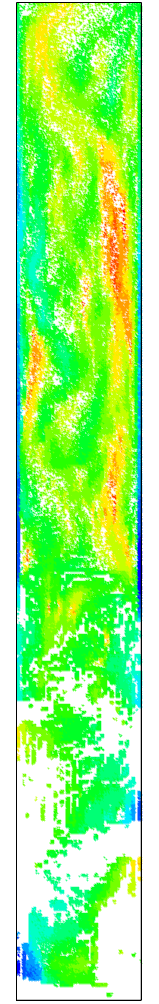}}
	\qquad		
	\subfigure{
		\includegraphics[height=8.0cm]{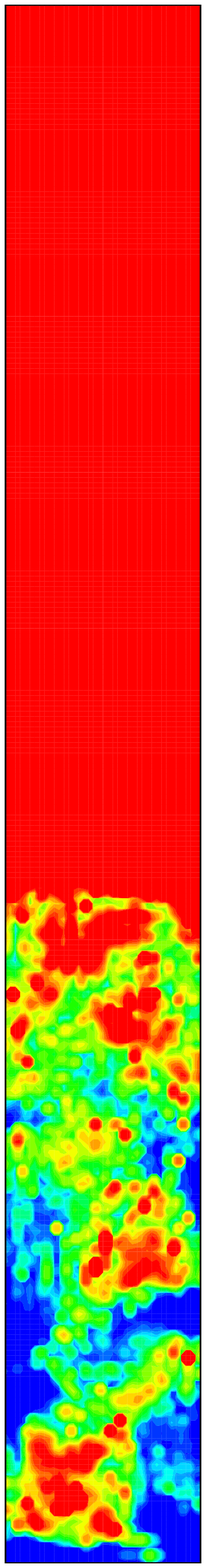}}
	\hspace{-5mm}
	\subfigure{
		\includegraphics[height=8.0cm]{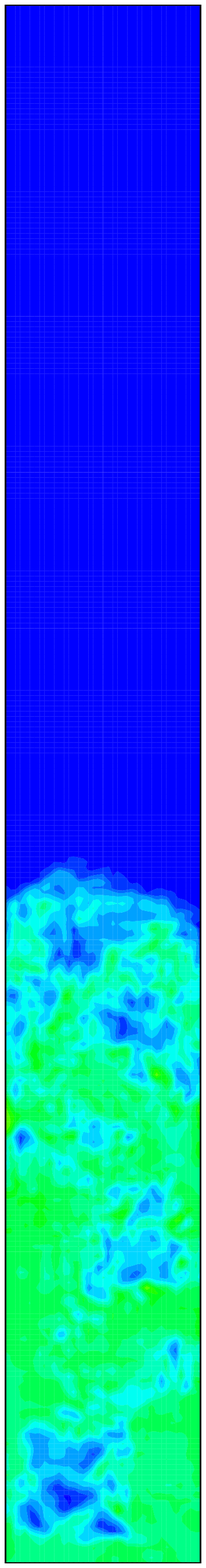}}
	\hspace{-5mm}
	\subfigure{
		\includegraphics[height=8.0cm]{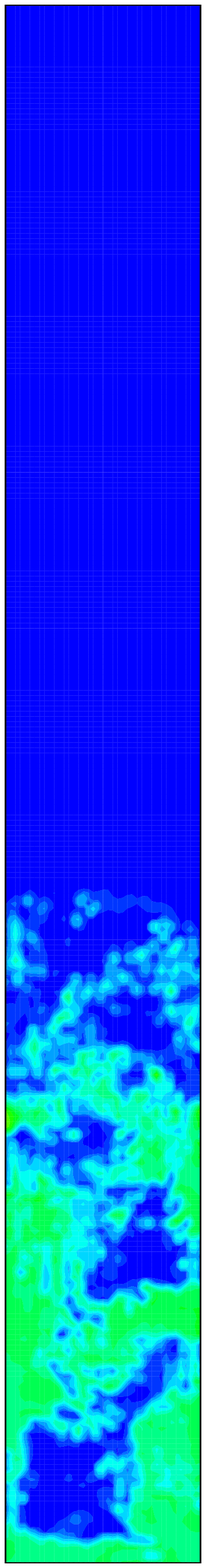}}
	\hspace{-1.5mm}
	\subfigure{
		\includegraphics[height=8.0cm]{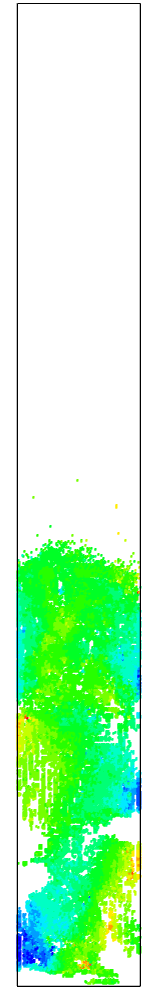}}
	\begin{center}
		\vspace{-2pt}		
		\footnotesize \quad~ $\text{Kn}_1$ \qquad \quad $\epsilon_{1}\rho_1$ \qquad \quad $\epsilon_{1}^{wave}\rho_1$ \qquad \quad $P_1$ \qquad \qquad \qquad $\text{Kn}_2$ \qquad \quad $\epsilon_{2}\rho_2$ \qquad \quad $\epsilon_{2}^{wave}\rho_2$ \qquad \quad $P_2$ \quad
	\end{center}	
	\caption{The instantaneous snapshots of $\text{Kn}_k$, solid apparent density $\epsilon_k\rho_k$, solid apparent density by wave in UGKWP $\epsilon^{wave}_k\rho_k$, and the set of sampled particles in UGKWP $P_k$ at $t=11.0s$. The subscript 1 and 2 stand for $1$st (FCC catalyst particle) solid phase and $2$nd (millet particle) solid phase, respectively. The $\text{Kn}_k$ is colored by the Kn-legend, the solid apparent density $\epsilon_k\rho_k$ and wave component $\epsilon^{wave}_k\rho_k$ are colored by the legend of apparent density, and the discrete particles in particle set $P_k$ are colored by the \text{Us}-legend (vertical velocity of the solid particle), with $k=1,2$. The legend of \text{Kn} is in the exponential distribution.}
	\label{TFB t11}
\end{figure}

Since the diameters of two solid phases (60$\mu m$ and 930 $\mu m$) have a difference of $15$ times,
it is interesting to present their respective distributions.
Figure \ref{TFB t11} shows the apparent density of each phase $\epsilon_{k}\rho_{k}$, $k=1,2$ at $t=11.0s$.
The summation $\sum_{k}\epsilon_{k}\rho_{k}$ is exactly the apparent density at $11.0s$ in Figure \ref{TFB instantaneous rhos}(b).
The obvious characteristic feature is that in the up dilute regions only the smaller (FCC catalyst) particles exist without millet particles,
as shown in the time-averaged profile of $\epsilon$ of Figure \ref{TFB time average eps rhos}.
Also, for both solid phases at $h\approx1.7m$, there is a separation zone with a large \text{Kn} above and a small \text{Kn} below.
Compared with the FFC catalyst phase, the millet phase shows a strong non-equilibrium with a large \text{Kn} in the bottom dense region.
The decompositions of wave and particle, determined by local \text{Kn}, are presented in Figure \ref{TFB t11} through
the contoured apparent density $\epsilon_{k}\rho_{k}$ and scattered particle set $P_k$.
For the FCC catalyst phase, the Lagrangian particle fully determines its evolution in the up-dilute region,
while  the wave component is dominant in the bottom region with tremendous amount of real particles and their collisions.
For the millet large particle phase, even in the bottom dense region, lots of particles are sampled and tracked in its evolution.

\begin{figure}[htbp]
	\centering
	\subfigure{
		\includegraphics[height=8.0cm]{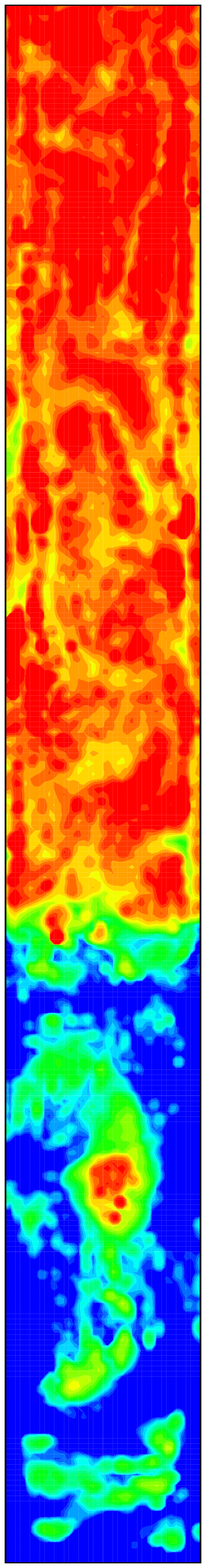}}
	\hspace{-8mm}
	\subfigure{
		\includegraphics[height=8.0cm]{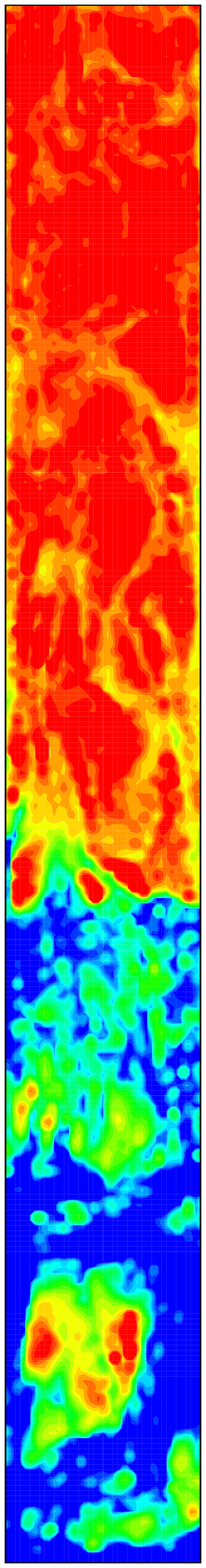}}
	\hspace{-8mm}
	\subfigure{
		\includegraphics[height=8.0cm]{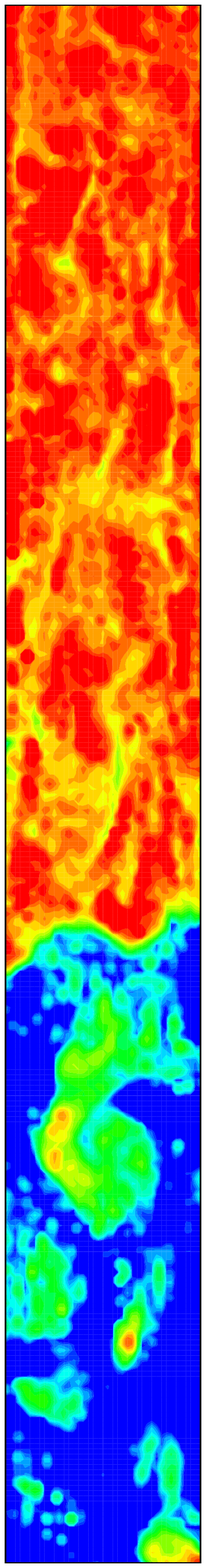}}
	\hspace{-8mm}	
	\subfigure{
		\includegraphics[height=8.0cm]{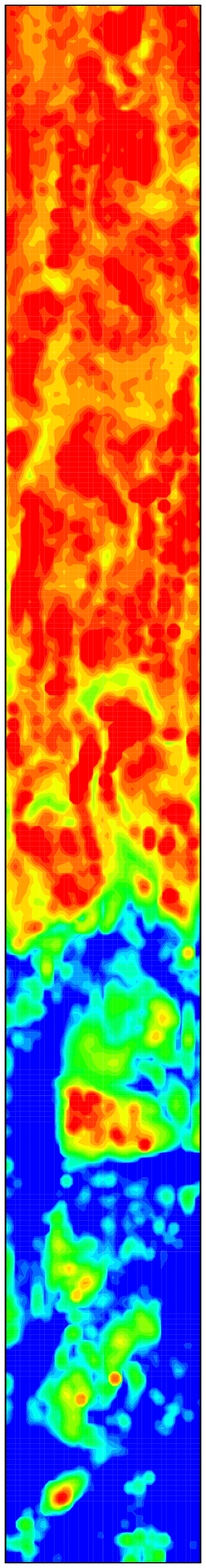}}
	\subfigure{
		\includegraphics[height=8.0cm]{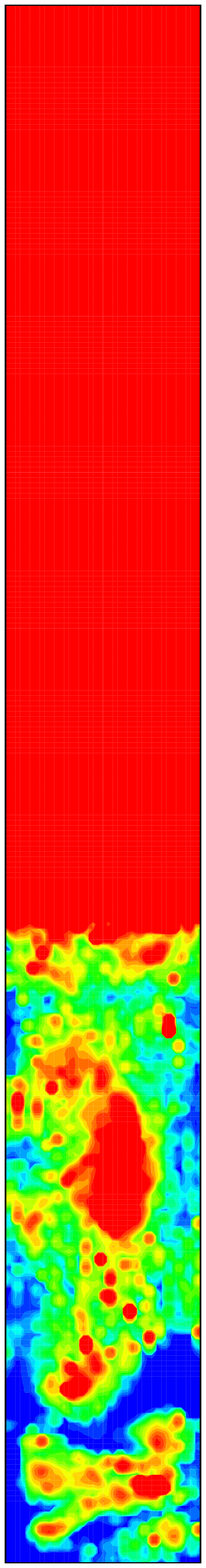}}
	\hspace{-8mm}
	\subfigure{
		\includegraphics[height=8.0cm]{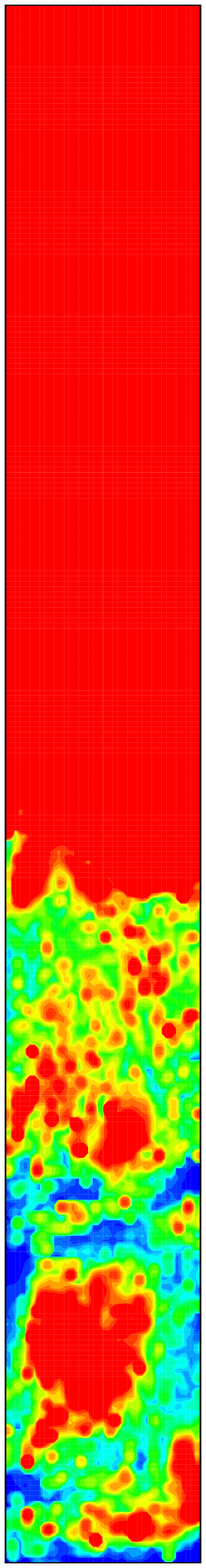}}
	\hspace{-8mm}
	\subfigure{
		\includegraphics[height=8.0cm]{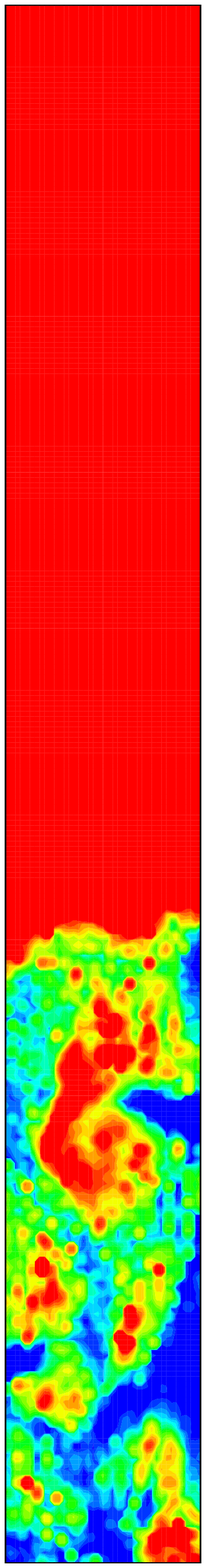}}
	\hspace{-8mm}	
	\subfigure{
		\includegraphics[height=8.0cm]{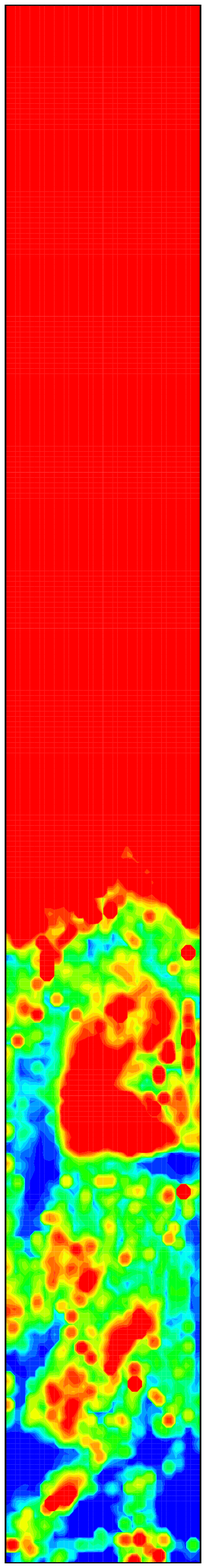}}
	\subfigure{
		\includegraphics[height=7.0cm]{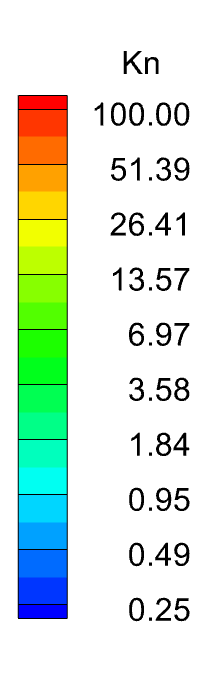}}
	\begin{center}
		\vspace{-2pt}		
		\footnotesize FCC catalyst phase \qquad\qquad\qquad\qquad\qquad\qquad\quad
		 Millet phase \qquad\qquad\qquad\qquad\qquad\qquad
	\end{center}	
	\caption{The instantaneous snapshots of \text{Kn} of PCC catalyst particle and millet particle phase at different times: from left to right are $t=10s, 12s, 14s, 15s.$ The legend of \text{Kn} is in the exponential distribution.}
	\label{TFB instantaneous Kn}
\end{figure}

It is important to emphasize the concepts of dilute/dense and non-equilibrium/equilibrium in gas-particle system.
The dilute or dense flow is generally determined by the solid volume fraction,
while non-equilibrium/equilibrium is determined by the \text{Kn} of the solid particle phase.
The dilute and dense flow regions can be associated with either non-equilibrium and equilibrium regimes, especially for the dense particle flow with much differences in their particle diameters or material density.
In UGKWP, \text{Kn} is used to determine the decomposition of wave and particle according to the extent of local non-equilibrium.
The snapshots of \text{Kn} for each disperse phase at different times are shown in Figure \ref{TFB instantaneous Kn}.
For both disperse phases, in the bottom dense region at $h<1.7m$ the non-equilibrium/equilibrium regimes are not spatially fixed and become dynamically inter-convertible. The proposed GKS-UGKWP  for polydisperse flow is suitable to treat individual solid phase with the optimal
 decompositions of wave and particle.

\subsection{Interaction of a shock wave with dense particle curtain}
\subsubsection{Case description}
\begin{figure}[htbp]
	\centering
	\subfigure[]{
		\includegraphics[height=5.0cm]{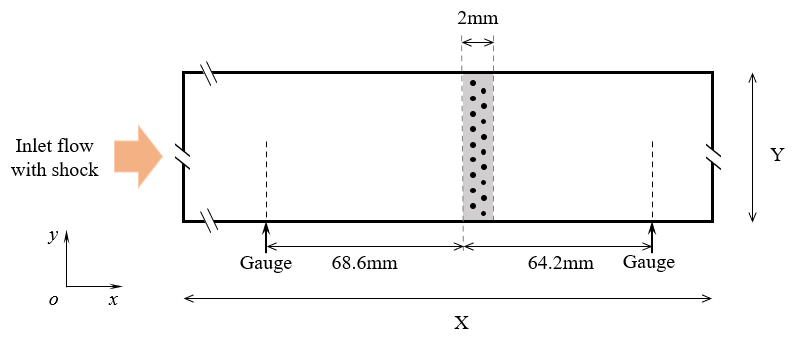}}
	\quad
	\subfigure[]{
		\includegraphics[height=4.5cm]{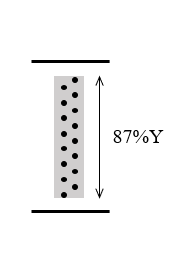}}
	\caption{The sketch of the interaction of a shock $Ma=1.66$ with the dense particle curtain: (a) the whole domain without near-wall gap; (b) the local particle distribution with near-wall gap (87\% spanwise particle curtain).}
	\label{sketch-shock-part}
\end{figure}

The interaction of a shock wave with solid particle bed is a highly challenging problem for a numerical method to be capable of 
capturing the shock wave propagation at supersonic speed and calculating the gas-solid phase-interaction, and particle-particle collisions at moderate/dense cases \cite{Gasparticle-shock-particle-curtain-ling2012interaction, Gasparticle-PIC-DEM-tian2020compressible}. 
Here, the interaction of a planar shock with a particle curtain is studied by the GKS-UGKWP method in two-dimensional space, and the simulation solution is compared with the experimental measurement \cite{Gasparticle-shock-particle-curtain-ling2012interaction}.
As sketched in Figure \ref{sketch-shock-part}(a), a planar shock with $\text{Ma}=1.66$ in the gas tube moves from the left to the right ($x$ direction), and encounters an initially stationary particle curtain with a width of $L=2mm$. 
Starting from the impingement of the shock on the solid particles bed, a reflecting shock moving to the left and a transmitting shock moving 
to the right will occur. Simultaneously, driven by the high-speed gas flow, the solid particles will move to the right by following the transmitted shock front.
The computational domain $X\times Y$ is $\left[-0.375m, 0.125m\right] \times \left[-0.04m, 0.04m\right]$ is covered by a uniform rectangular mesh with $1000\times160$ cells.
In the experiment, the diameter of solid particles is distributed by $106\mu m\sim 125\mu m$. 
While in the computation, the simulation is based on two disperse phases with solid particle diameters $d_1=110\mu m$ and $d_2=120\mu m$. 
The material density of all solid particles is $\rho_s = 2520kg/m^3$. 
Initially, the solid particles are uniformly distributed with $\epsilon_{1}=0.105$ and $\epsilon_{2}=0.105$.
The initial state of the gas phase in the domain is the same as the experiment: $p_{g,1}=8.27\times10^{4}Pa$, $U_{g,1}=0m/s$ and $T_{g,1}=296.40K$, 
with gas constant $R=287.05 J/(kg\cdot K)$ and specific heat ratio $\gamma=1.4$. 
At the left boundary, the pre-shock condition at $\text{Ma}=1.66$ is given by $p_{g,2}=2.52\times10^{5}Pa$, $U_{g,2}=304.16m/s$ and $T_{g,2}=423.79K$. 
The free boundary condition is employed at the right boundary. 
Besides, for the top and bottom boundaries, the non-slip and slip boundary condition are taken for the gas and solid particle phases, respectively.
To monitor the pressure of the gas flow, two gauge positions are set at $68.6mm$ upstream and $64.2mm$ downstream of the left front of the initial particle curtain.
In this case, the internal degree of freedom of the gas phase is modeled by $K\left(\tilde{t}\right) = K_0 + 1.0\times\left(\tilde{t}/\tilde{t}_{end}\right)^3$ to mimic the increased turbulence intensity of gas flow due to the interaction with disperse solid particles. 
$\tilde{t}=\frac{t}{L/u_s}$ is the normalized time by the initial width of particle curtain $L$ and shock velocity $u_s=572.00m/s$, and $\tilde{t}_{end}$ is the normalized simulation time taken as $350.0$ in this case.

The drag force on the solid particle of $k$-th disperse phase can be written in a general form,
\begin{gather*}
\textbf{D} = 3\pi \mu_g d_k \left(\textbf{U}_g-\textbf{u}\right) \frac{Re_k}{24} C_D^*,
\end{gather*}
where $Re_k$ is the particle Reynolds number, and $C_D^*$ is the particle drag coefficient. 
With the definition of $\textbf{D}$ in Eq.\eqref{drag force model}, $\tau_{st,k}$ can be obtained,
\begin{gather}
\tau_{st,k} = \frac{4}{3} \frac{\rho_k d_k^2}{\mu_g C_D^* Re_k}.
\end{gather}
In this paper, the particle drag coefficient $C_D^*$ is calculated by
$C_D^* = c_1\left(\epsilon_{k}\right) c_2 C_{D,tad}$,
where $C_{D,std}$ is the standard drag correlation proposed by Clift \cite{Gasparticle-drag-cd-clift1970motion},
\begin{gather*}
C_{D,std} = \frac{24}{Re_k} \left(1.0 + 0.15 Re_k^{0.687}\right)
+ 0.42 \left(1.0 + \frac{42500}{Re_k^{1.16}}\right),
\end{gather*}
$c_1\left(\epsilon_{k}\right) = \frac{1 + 2 \epsilon_{k}}{\left(1 - \epsilon_{k}\right)^2}$ is the correlation factor for the effect of the finite particle volume fraction given by Sangani et al. \cite{Gasparticle-drag-cd-cor-parmar2008unsteady},
and $c_2$ is the correlation factor with a value $4.0$ for a better agreement with the experimental measurement, 
which can be interpreted as the collective effect from other forces, such as added-mass force, viscous-unsteady force, etc.

\subsubsection{Results}
\begin{figure}[htbp]
	\centering
	\subfigure[]{
		\includegraphics[height=6.5cm]{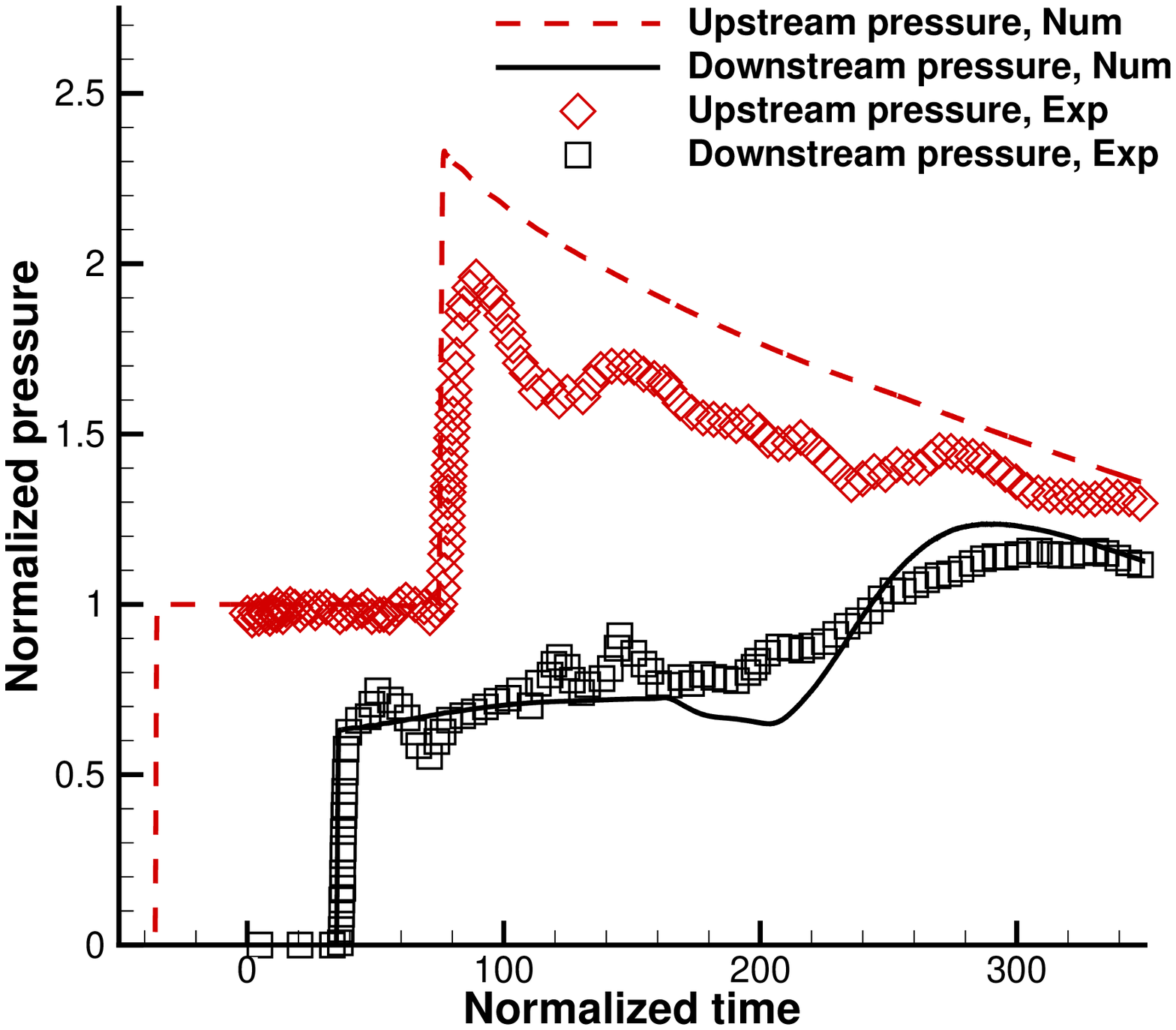}}
	\quad
	\subfigure[]{
		\includegraphics[height=6.5cm]{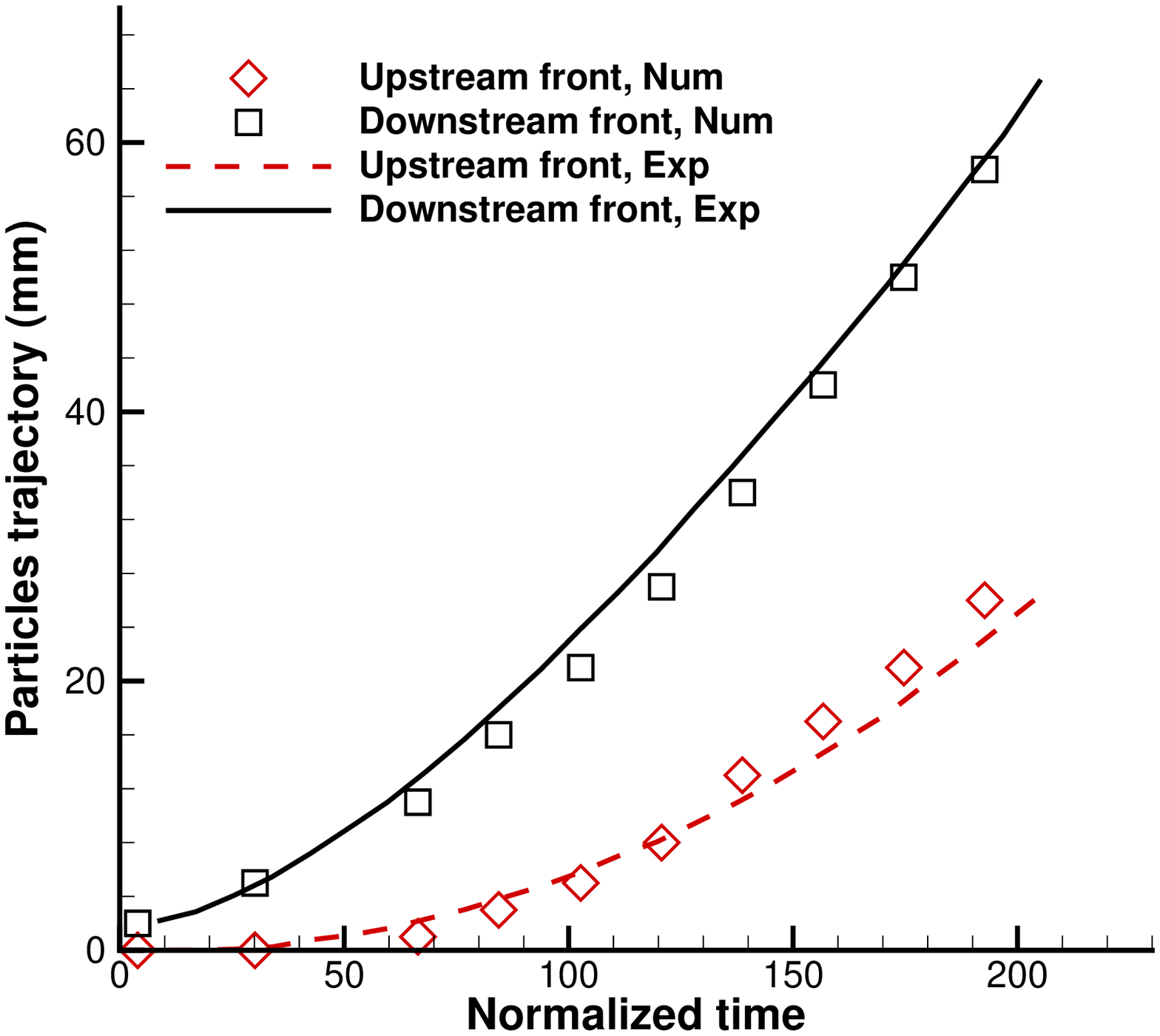}}
	\caption{Numerical results (denoted by ``Num") and experimental measurements (denoted by ``Exp"): 
(a) the time-dependent gas pressure at upstream and downstream gauge points; (b) the trajectories of particle cloud fronts.}
	\label{shockpart p and x}
\end{figure}

\begin{figure}[htbp]
	\centering	
	\subfigure[All the solid particles]{
		\includegraphics[height=6.5cm]{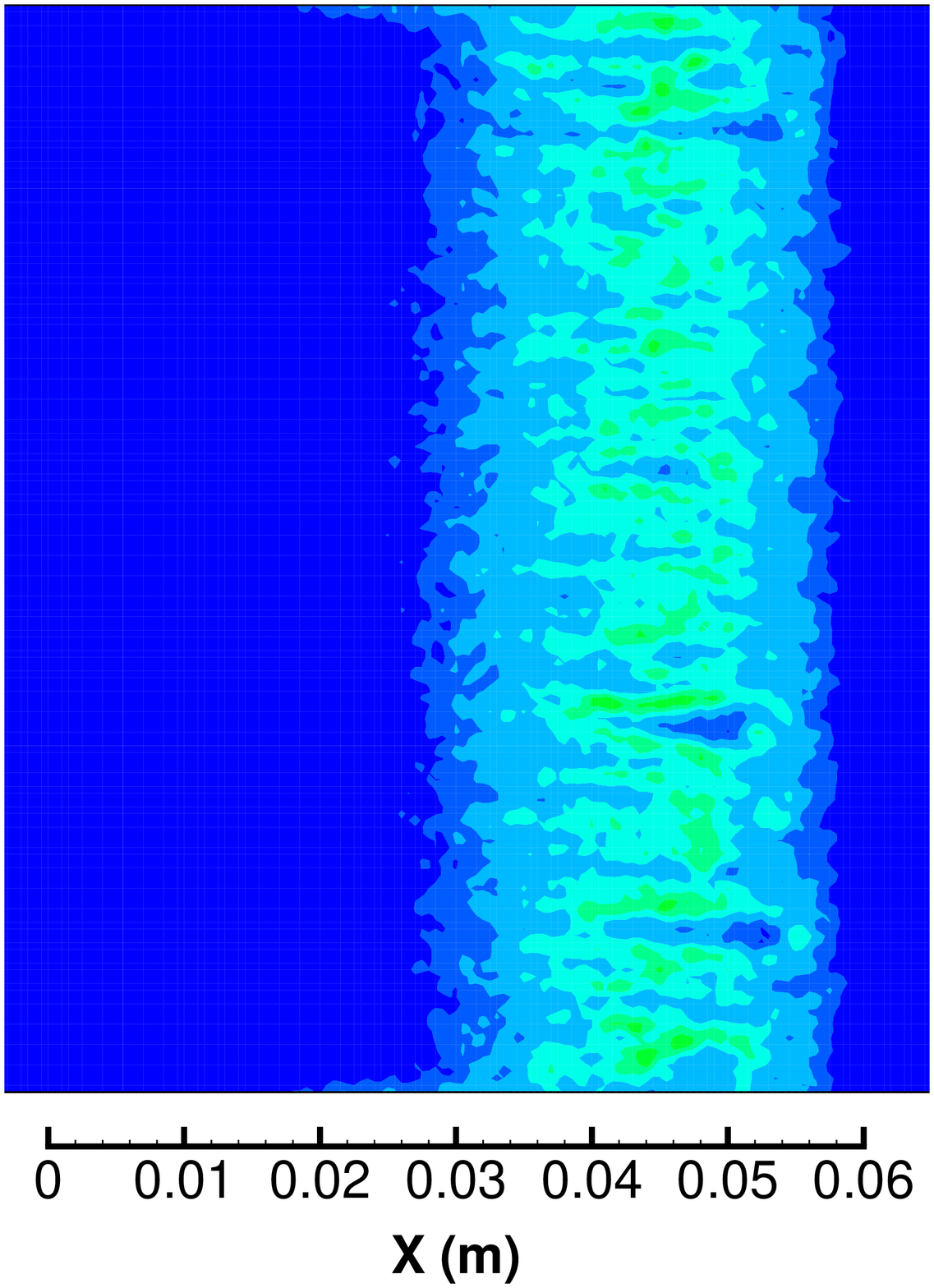}}	
	\quad
	\subfigure[Solid particles with $d_1$]{
		\includegraphics[height=6.5cm]{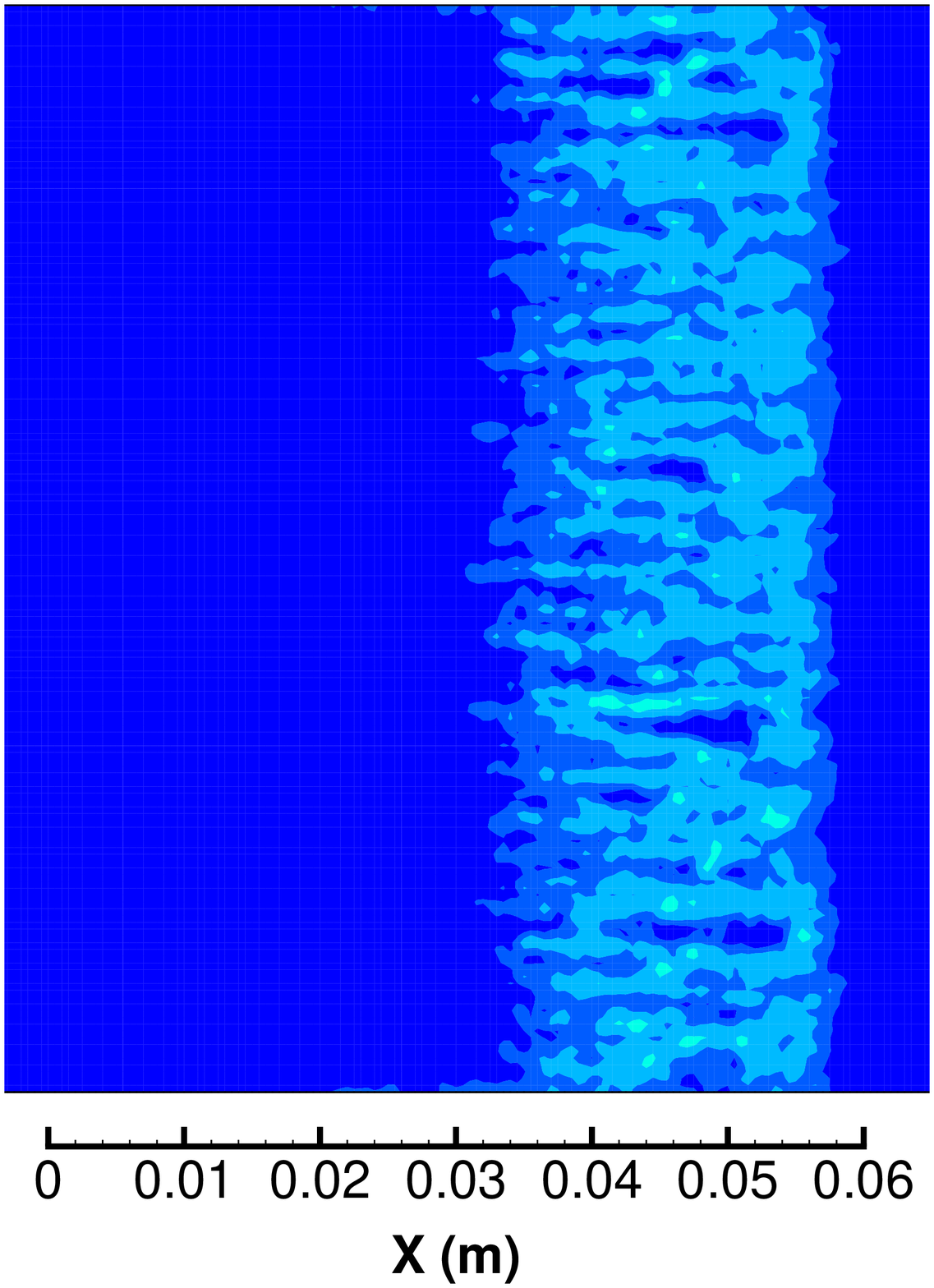}}
	\quad
	\subfigure[Solid particles with $d_2$]{
		\includegraphics[height=6.5cm]{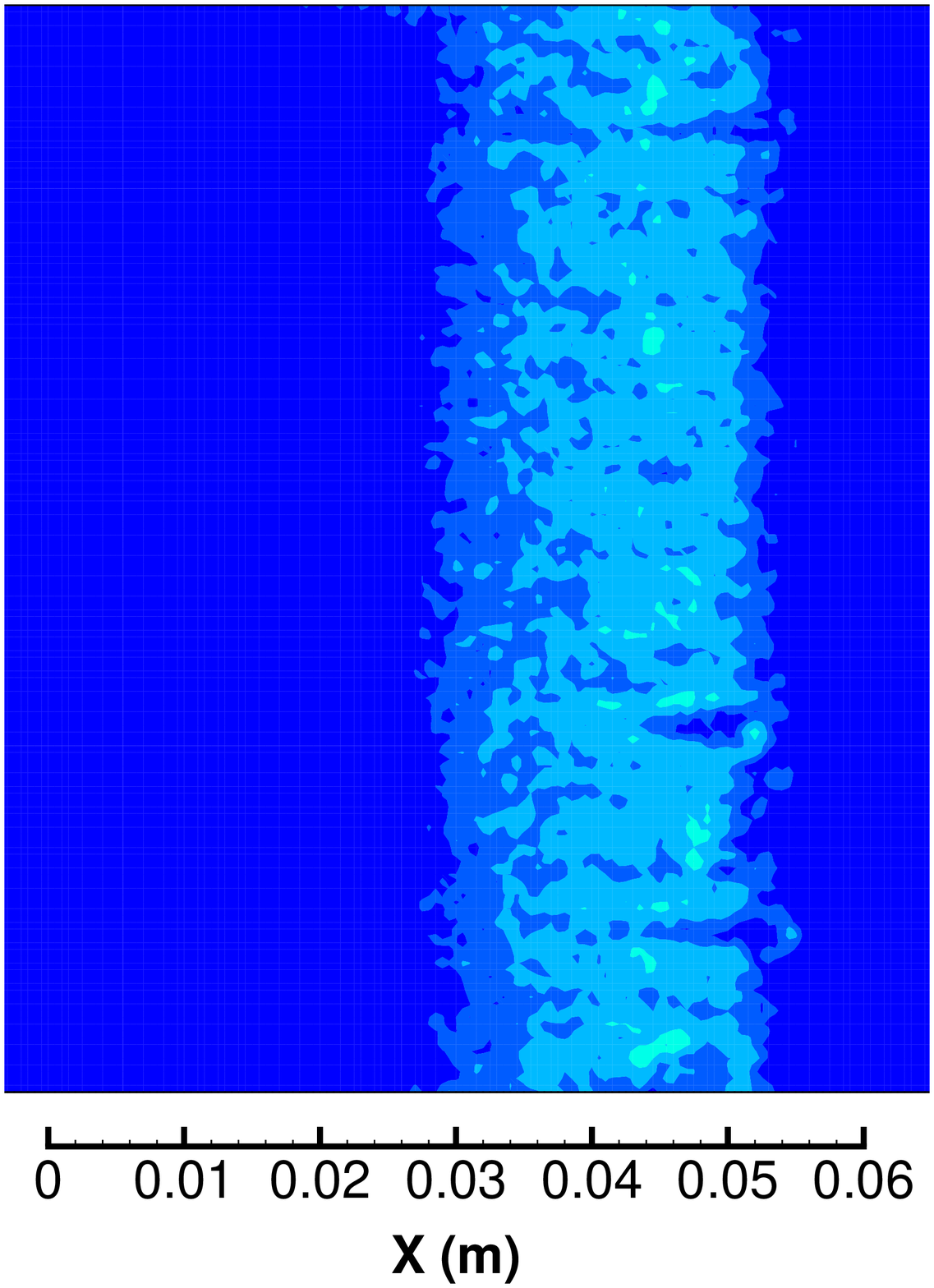}}
	\quad
	\subfigure{
		\includegraphics[height=5.0cm]{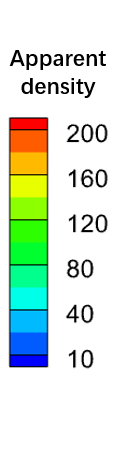}}
	\caption{Instantaneous snapshots of the apparent density of particle phases at a normalized time $\tilde{t}=193.0$: (a) the whole solid particle phases, (b) the 1st particle phase with $d_1=110\mu m$, and (c) the 2nd particle phase with $d_2=120\mu m$. 
Note that only the region of $\left[0m, 0.06m\right] \times \left[-0.04m, 0.04m\right]$ is shown here.}
	\label{shockpart rho nortime 193}
\end{figure}

The time-dependent pressure at gauging positions is presented in Figure \ref{shockpart p and x}(a) 
and compared with the experimental measurements.
Note that the pressure at upstream and downstream gauging positions are averaged in $y$ direction. 
As used in \cite{Gasparticle-shock-particle-curtain-ling2012interaction}, the pressure $p_g\left(\tilde{t}\right)$ is normalized by $\frac{p_g\left(\tilde{t}\right) - p_{g,1}}{p_{g,2} - p_{g,1}}$. 
In general, both the reflected shock and transmitted shock can be captured well.
Besides, the trajectories of the upstream front and downstream front of the solid particle cloud are shown in Figure \ref{shockpart p and x}(b), which agree well with the experiment data.
The instantaneous snapshot of the distribution of solid particles at $\tilde{t}=193.0$ is presented in Figure \ref{shockpart rho nortime 193}. 
The solid particles are not uniformly distributed in the existing region: the central zone (around $x=0.045m$) shows higher concentration than the regions near the upstream and downstream fronts. Also, a slight particle-cluster phenomenon can be observed. 
Further comparison of 1st and 2nd dispersed phases shows that the 1st solid phase with smaller particle $d_1=110\mu m$ moves faster (approximately $3mm$) than the 2nd phase with $d_2=120\mu m$ in both upstream front and downstream fronts.

\begin{figure}[htbp]
	\centering
	\subfigure{
		\includegraphics[height=6.5cm]{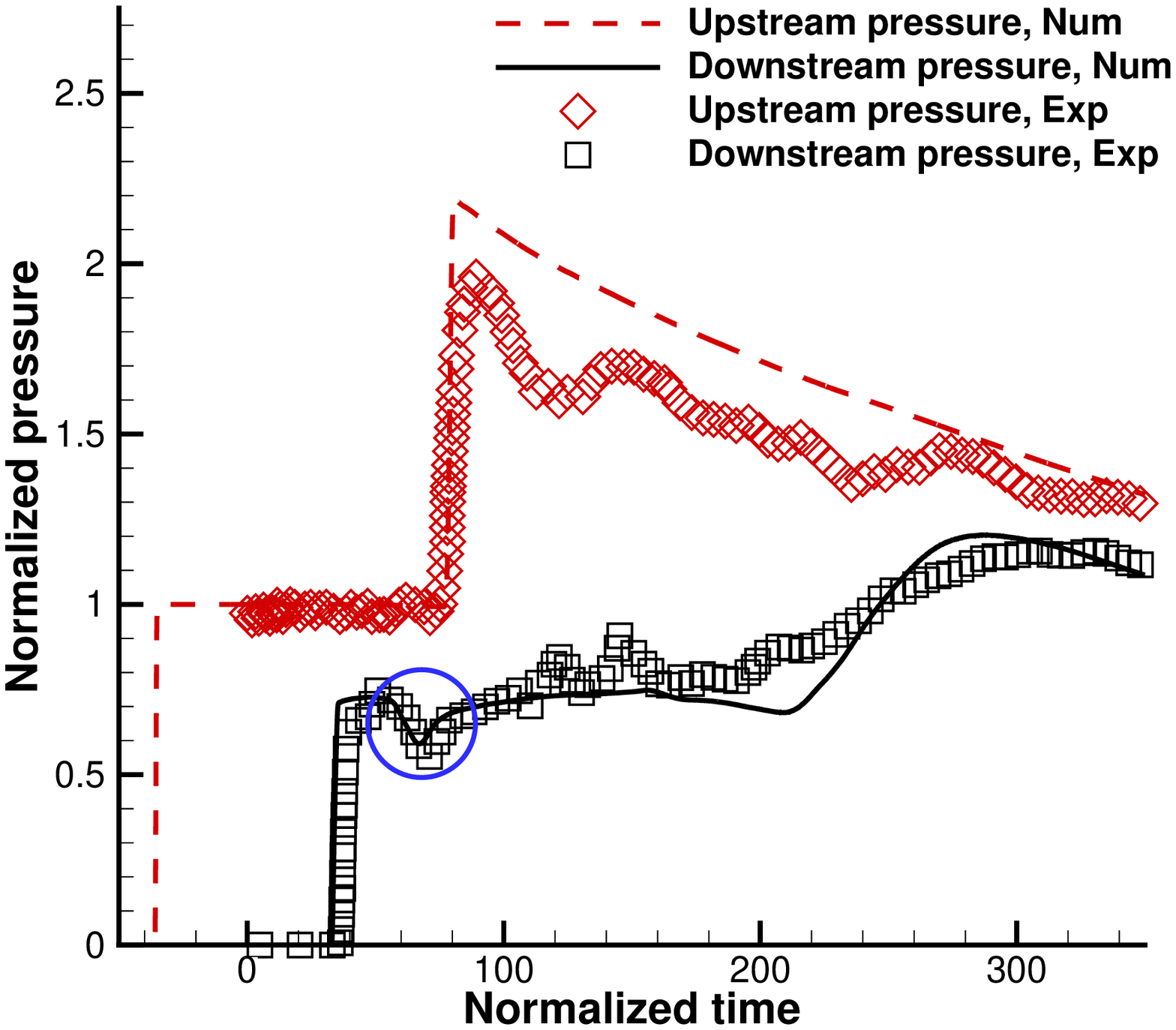}}
	\quad
	\subfigure{
		\includegraphics[height=6.5cm]{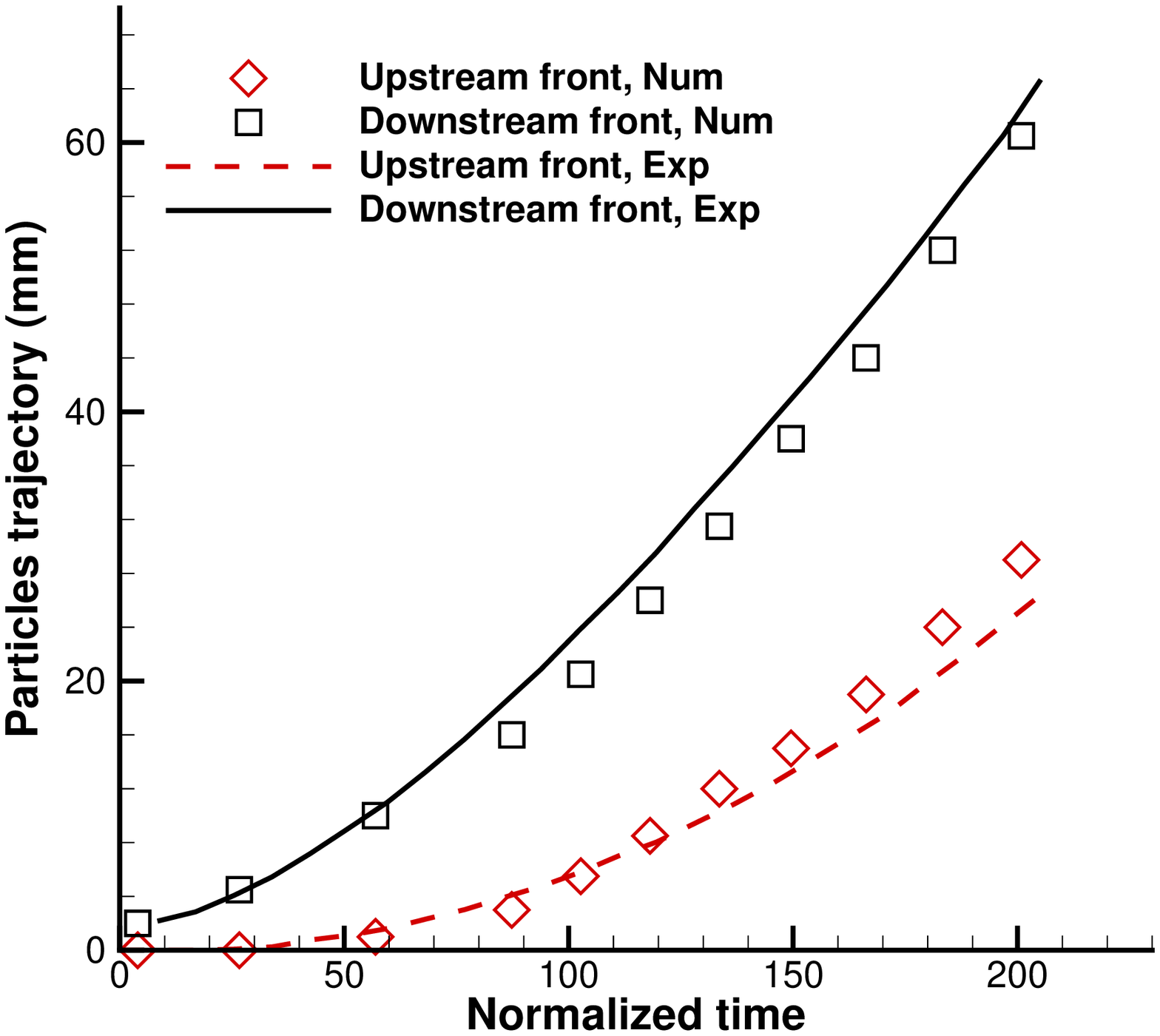}}
	\caption{Numerical results (denoted by ``Num") and experimental measurements (denoted by ``Exp") with $87\%$ span-wise particle occupation as shown Figure \ref{sketch-shock-part}(b): (a) the time-dependent gas pressure at upstream and downstream gauge points; (b) the trajectories of particle cloud fronts.}
	\label{shockpart p and x gap}
\end{figure}

In the experiment, the solid particle curtain is generated by the free fall of particles from a  reservoir into the test section \cite{Gasparticle-shock-particle-curtain-ling2012interaction}. 
As pointed out in the experiment, the particle curtain occupies about 87\% in the span-wise direction ($y$ direction in Figure \ref{sketch-shock-part}). Therefore, a gap between the particle curtain and the walls exist, which is studied here as well. 
According to the experiment with 87\% occupation by particle curtain in the tube, a 13\% gap close to the wall will be taken into account. 
With the new set-up, the newly calculated pressure and particles' trajectories are given in Figure \ref{shockpart p and x gap}. 
Interestingly, the pressure variation at the downstream gauge position at time $\tilde{t}=50\sim80$ has an early  
drop and a later increase in gas pressure after the passage of the transmitted shock and has a better agreement with the experimental data than the previous calculation with 100\% particle curtain occupation, which is marked by the blue circle in Figure \ref{shockpart p and x gap}.

\section{Conclusion}
In this paper, a multiscale GKS-UGKWP method is developed for polydisperse gas-particle system.
Particularly, the cell resolution dependent kinetic model for the system with coupled disperse solid particle phase and gas phase 
is constructed and used in the design of the corresponding multiscale method. 
In order to capture both equilibrium and non-equilibrium states in the particle phase evolution, 
the particle distribution is decomposed into wave and discrete particle components in UGKWP
according to the respective local \text{Kn} for each solid particle phase.
The UGKWP will automatically choose the optimal way to describe the solid particle dynamics by the combination of  
the deterministic wave and statistical particle with the full consideration of physical accuracy and numerical efficiency, 
for the individual solid particle phase with different physical properties, such as particle size, concentration, and material.
One distinguishable feature in UGKWP is that the wave-particle decomposition can automatically recover the Eulerian-Eulerian and Eulerian-Lagrangian methods for the gas-particle system in the corresponding particle continuum and non-equilibrium regimes. 
Two cases of dense polydisperse flow in the fluidization bed are simulated. 
Specifically, in the FCC catalyst reactor, the diameters of large millet particle and small FCC catalyst particle differ in particle size by $15$ times. The numerical experiments show that only the fine FCC catalyst particles appear in the top zone of the riser, while the millet particles only exist in the bottom dense region.
In addition, for the FCC catalyst phase, the discrete particle description plays key role in the top-dilute regions, 
and the wave description contributes more in the bottom-dense zone region even with the existence of tremendous amount of real particles.
For the large particle phase, millet particles only exist in the bottom dense region and need discrete particle description to 
capture local non-equilibrium state. 
The above observation indicates the flexibility of wave-particle decomposition in describing the disperse solid phases and 
the adaptivity in dynamically following the equilibrium and non-equilibrium flow evolution.
At the same time, the interaction of a $Ma=1.66$ shock wave impinging on a dense particle curtain with polydisperse solid particle phases is studied by the current method. The numerical solutions, such as the pressures of the gas flow at gauge points and the trajectories 
of the fronts of solid particle cloud, are compared with the experimental measurements. 
Overall, the performance of the GKS-UGKWP is favorable in comparison with the existing single scale methods for the 
simulation of gas-particle multiphase flow. 

\section*{Acknowledgements}
The current research is supported by National Key R\&D Program of China (Grant Nos. 2022YFA1004500), National
Science Foundation of China (12172316), and Hong Kong research grant council (16208021, 16301222).

\bibliographystyle{plain}%
\bibliography{reference}
\end{document}